\documentclass[namedreferences]{solarphysics}
\usepackage[hyperref,optionalrh,solaromanenum]{spr-sola-addons} 
\usepackage{graphicx}                    
\usepackage{color}                       
\usepackage[flushleft]{threeparttable}

\newcommand{\aap}{    {\it Astron. Astrophys.}}
\newcommand{\aaps}{   {\it Astron. Astrophys. Suppl.}}

\newcommand{\apj}{    {\it Astrophys. J.}}

\newcommand{\apjl}{   {\it Astrophys. J. Lett.}}

\newcommand{\jgr}{    {\it J. Geophys. Res.}}
\newcommand{\mnras}{  {\it Mon. Not. Roy. Astron. Soc.}}

\newcommand{\solphys}{{\it Solar Phys.}}
 
\newcommand{\ssr}{    {\it Space Sci. Rev.}} 
\chardef\us=`\_

\begin{document}

\begin{article}

\begin{opening}

\title{Statistical Analysis and Catalog of Non--polar Coronal Holes Covering the SDO--era using CATCH}

%
\author[addressref={graz},corref,email={stephan.heinemann@hmail.at}]{\inits{S.G.}\fnm{Stephan G. }\lnm{Heinemann}\orcid{0000-0002-2655-2108}}
\author[addressref={graz},corref,email={}]{\inits{M.}\fnm{Manuela }\lnm{Temmer}\orcid{0000-0003-4867-7558}}
\author[addressref={graz},corref,email={}]{\inits{N.}\fnm{Niko }\lnm{Heinemann}\orcid{}}
\author[addressref={graz},corref,email={}]{\inits{K.}\fnm{Karin }\lnm{Dissauer}\orcid{0000-0001-5661-9759}}
\author[addressref={leuven,rob},corref,email={}]{\inits{E.}\fnm{Evangelia }\lnm{Samara}\orcid{}}
\author[addressref={graz,zagreb},corref,email={}]{\inits{V.}\fnm{Veronika }\lnm{ Jer{\v c}i{\' c}}\orcid{}}
\author[addressref={graz},corref,email={}]{\inits{S. J.}\fnm{Stefan J. }\lnm{Hofmeister}\orcid{0000-0001-7662-1960}}
\author[addressref={graz,kso},corref,email={}]{\inits{A.}\fnm{Astrid M. }\lnm{Veronig}\orcid{0000-0003-2073-002X}}

%
\runningauthor{S.G. Heinemann \textit{et al.}}
\runningtitle{CATCH}

\address[id={graz}]{University of Graz, Institute of Physics, Universit\"atsplatz 5, 8010 Graz, Austria }
\address[id={leuven}]{Centre of Mathematical Plasma Astrophysics, KU Leuven, Leuven, Belgium}
\address[id={rob}]{Royal Observatory of Belgium, Brussels, Belgium}
\address[id={zagreb}]{University of Zagreb, Faculty of Science, Department of Geophysics, Zagreb, Croatia}
\address[id={kso}]{Kanzelh\"ohe Observatory for Solar and Environmental Research, University of Graz, 9521 Treffen, Austria}

\begin{abstract}
Coronal holes are usually defined as dark structures as seen in the extreme ultraviolet and X-ray spectrum which are generally associated with open magnetic field. Deriving reliably the coronal hole boundary is of high interest, as its area, underlying magnetic field, and other properties give important hints towards high speed solar wind acceleration processes and on compression regions arriving at Earth. In this study we present a new threshold based extraction method that incorporates the intensity gradient along the coronal hole boundary, which is implemented as a user-friendly SSWIDL GUI. The Collection of Analysis Tools for Coronal Holes (CATCH) enables the user to download data, perform guided coronal hole extraction and analyze the underlying photospheric magnetic field. We use CATCH to analyze non-polar coronal holes during the SDO-era, based on $193$ \AA\ filtergrams taken by the \textit{Atmospheric Imaging Assembly} (AIA) and magnetograms taken by the \textit{Heliospheric and Magnetic Imager} (HMI), both on board the \textit{Solar Dynamics Observatory} (SDO). Between $2010$ and $2019$ we investigate $707$ coronal holes that are located close to the central meridian. We find coronal holes distributed across latitudes of about $\pm 60^{\mathrm{o}}$ for which we derive sizes between $1.6 \times 10^{9}$ and $1.8 \times 10^{11}$ km$^{2}$. The absolute value of the mean signed magnetic field strength tends towards an average of $2.9\pm 1.9$ G. As far as the abundance and size of coronal holes is concerned, we find no distinct trend towards the northern or southern hemisphere. We find that variations in local and global conditions may significantly change the threshold needed for reliable coronal hole extraction and thus, we can  highlight the importance of individually assessing and extracting coronal holes.
\end{abstract}

%

\end{opening}

\section{Introduction}\label{s:intro}
Coronal holes (CHs) are large-scale features in the solar corona often characterized by reduced emission in X-ray and extreme ultraviolet (EUV) which are associated with open magnetic field lines of a dominant polarity. Coronal plasma is accelerated along the open field lines causing a high velocity outflow of particles, often referred to as fast solar wind or high speed solar wind stream (HSS). The plasma depletion causes a reduction of density and temperature in these regions in comparison to the surrounding solar corona. Thus, CHs can be observed as dark structures in the EUV and X-ray emission \citep[see \textit{e.g.,}][]{schwenn06,cranmer2002,cranmer2009}. 

To investigate the morphology and intensity of CHs as observed in EUV, as well as their underlying photospheric magnetic field, the identification and extraction of CH boundaries are key. There exist multiple approaches to this topic with one of the most popular using a single wavelength, intensity-based threshold approach on EUV observations. Due to the high contrast and the optimal filter sensitivity, the coronal emission line of eleven times ionized iron (Fe \textsc{xii}: $193/195$ \AA) is often used to extract CHs \citep[\textit{e.g.,}][]{2009krista,2012rotter,2015rotter,2015reiss,2016caplan_LBC_IIT,Boucheron2016,2017hofmeister,2018heinemann_paperI}. Other intensity based approaches include multi-thermal emission recognition \citep{2018garton} and spatial possibilistic clustering \citep{2014spoca}.
A different concept is to model the open field, that characterizes CHs, using photospheric magnetograms. Examples include the potential field source surface model \citep[PFSS;][]{1969altschuler_PFSS}, its improved version including the Schatten Current Sheet, the Wang-Sheeley-Arge model \citep[WSA model;][]{2000arge} and the MULTI-VP model \citep{2017pinto}. Studies comparing the two different conceptual approaches have shown significant differences in the size, location, shape and occurrence of the dark and/or open structure defined as CHs \citep[\textit{e.g.,}][]{2014lowder,2017lowder,Linker_2017,2019wallace,2019huang, 2019asvestari}. Additionally new approaches like machine learning/neural networks \citep[\textit{e.g.,}][]{2018illarionov} and extraction methods based on plasma properties \citep[Differential Emission Measure;][]{1981raymond,2011hahn} are the topic of current research. 
 
Reliably defining CH boundaries is not only relevant for studying coronal and photospheric properties and their evolution but is also of major scientific importance towards space weather research. Empirical relations between CH area and measured solar wind speed at 1AU \citep[\textit{e.g.,}][]{1976nolte,2007vrsnak,2017tokumaru,2018hofmeister} are used for forecasting purposes \citep[\textit{e.g.,}][]{2012rotter,2015rotter,2016reiss,temmer18}. Moreover, the distance to CH boundary is an important parameter for MHD models simulating the solar wind distribution in interplanetary space (\textit{e.g.,} ENLIL: \citealt{1999enlil}, EUHFORIA: \citealt{2018euhforia} and CORHEL: \citealt{2012riley_CORHEL}). When considering CH extraction, usually there is the choice between manual and automated algorithms of which both have advantages and disadvantages. On the one hand, manual extraction of CHs requires a lot of time and experience in order to get reliable results. On the other hand, automated extraction methods are prone to significant errors and artifacts.\\
 
In the first part of this study we present a new method for extracting CH boundaries in EUV images by using an intensity threshold which is modulated by the intensity gradient of the CH boundary. The method is based on the works of \cite{2012rotter}, \cite{2015rotter}, and \cite{2009krista} and is incorporated into an easy-to-use GUI application developed in SSW-IDL. The Collection of Analysis Tools for Coronal Holes (CATCH) application enables users to easily extract and analyze CHs in a supervised semi-automated fashion. CATCH uses a modulated intensity threshold method to extract CH boundaries from EUV images and analyzes the associated properties. In addition, it offers the possibility to investigate the underlying magnetic field. In the second part, we use CATCH to investigate $707$ CHs covering the complete time range of the operational lifetime of the Solar Dynamics Observatory \citep[SDO;][]{2012pesnell_SDO} so far, starting in May $2010$ until February $2019$. We derive statistical CH properties of the area, intensity, and the underlying magnetic field including the magnetic fine structure over nearly the full Solar Cycle $24$. Furthermore, we present how the parameters for an optimal CH extraction vary during the Solar Cycle. The CH dataset is available as an online catalogue under the CDS database using the Vizier catalogue service \citep[][]{2000Ochsenbein_vizier}.

\section{The ``Collection of Analysis Tools for Coronal Holes"}\label{subs:catch}
 
The Collection of Analysis Tools for Coronal Holes (CATCH) was created in order to collect and structure CH identification, extraction and analysis in a handy and fast way without the disadvantages of automatic algorithms as described in the Sections~\ref{subs:euv} and \ref{subs:mag}. It enables the user to download and process EUV filtergrams ($193/195$ \AA) and line--of--sight (Los) magnetograms. CATCH is able to handle data from different spacecraft missions covering the interval from 1996 until now. These are SDO, the \textit{Solar Terrestrial Relations Observatory} \citep[STEREO;][]{2008kaiser_STEREO} and the \textit{Solar and Heliospheric Observatory} \citep[SOHO;][]{1995soho}. Data from the \textit{Atmospheric Imaging Assembly} \citep[AIA; $193$\AA]{2012lemen_AIA}, the \textit{Extreme ultraviolet Imaging Telescope} \citep[EIT; $195$\AA]{1995delaboudiniere-EIT} and the \textit{Extreme UltraViolet Imager} \citep[EUVI; $195$\AA]{2008howard_SECCHI} as well as from the \textit{Heliospheric and Magnetic Imager} \citep[HMI: ][]{2012schou_HMI,2016couvidat_HMI}  and the \textit{Michelson Doppler Imager} \citep[MDI: ][]{1995scherrer_mdi} can be processed. Additionally, user supplied full--disk images can also be analyzed.

\begin{figure} 
 \centerline{\includegraphics[width=1\textwidth,clip=]{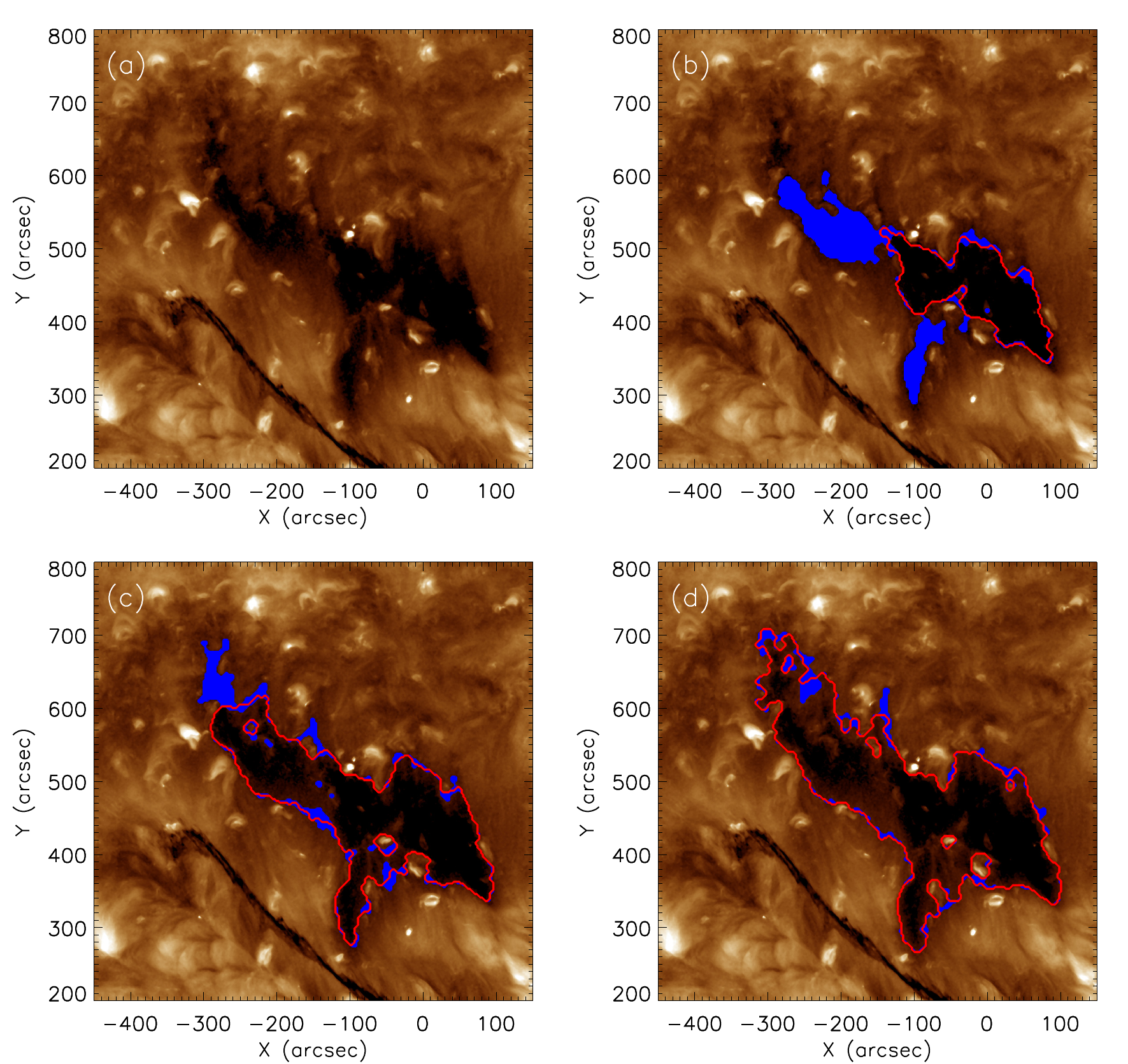}}
 \caption{Example images of boundary extraction of a CH on August 31, 2014. The red line is the CH boundary (of the chosen threshold)} and the blue shaded areas are the uncertainties as described in Section~\ref{subs:euv}. Panel (a) shows the CH without boundaries and panel (b) with the CH boundary of a threshold of $31$ DN which gives an uncertainty of $\epsilon_{A}= 47 \%$. In panel (c) a threshold of $37$ DN is used which gives an uncertainty of $\epsilon_{A}= 16\%$. The best boundary for this case (\textit{e.g.,} the lowest uncertainty) is reached with a threshold of $41$ DN and is shown in panel (d) with $\epsilon_{A}= 10\%$.\label{fig:example_boundary}
 \end{figure}

The user can perform CH boundary detection, extraction and analysis using a manually adjustable intensity threshold. The threshold range, in which reasonable CH boundaries can be extracted, can be derived from the intensity histogram of the solar disk. After specifying a threshold, it is applied to the full solar disk and the user may select the structure of interest to calculate its parameters and to get an estimate of the boundary stability and uncertainty. Then by varying the threshold to minimize the boundary uncertainty ($\epsilon_{A}$), the user can find an optimized CH boundary in an easy and fast way, even without previous experience in CH extraction. For deriving the properties of a CH, CATCH analyzes five boundaries in an interval of $1$ DN (data number) centered around the selected threshold and calculates the mean values. The maximum deviation of the derived values from the calculated mean is the uncertainty. After extracting a satisfactory boundary from EUV filtergrams, CATCH can analyze the properties of the CH. The boundary may then be used on LoS magnetograms (if available) to analyze the underlying photospheric magnetic field of the CH and its fine structure represented by FTs. Figure~\ref{fig:example_boundary} shows an example of how to find the optimal threshold by considering the uncertainty of the extracted CH boundary. The red contour represents the CH boundary (of the chosen threshold), the blue shaded areas are the uncertainties of the boundary. The best boundary for this CH can be identified as shown in panel (d), where the blue shaded area is smallest in comparison to the area enclosed in the CH boundary.
 
CATCH calculates a variety of properties of the extracted CH, which include morphological properties, the intensity, boundary stability as well as properties of the underlying photospheric magnetic field and its fine structure (for the full list of calculated parameters see Tab.~\ref{tab:params}). The calculations are based on the studies by \cite{2017hofmeister}, \cite{2018heinemann_paperI}, and \cite{2018heinemann_paperII}.

\begin{figure} 
 \centerline{\includegraphics[width=1\textwidth,clip=]{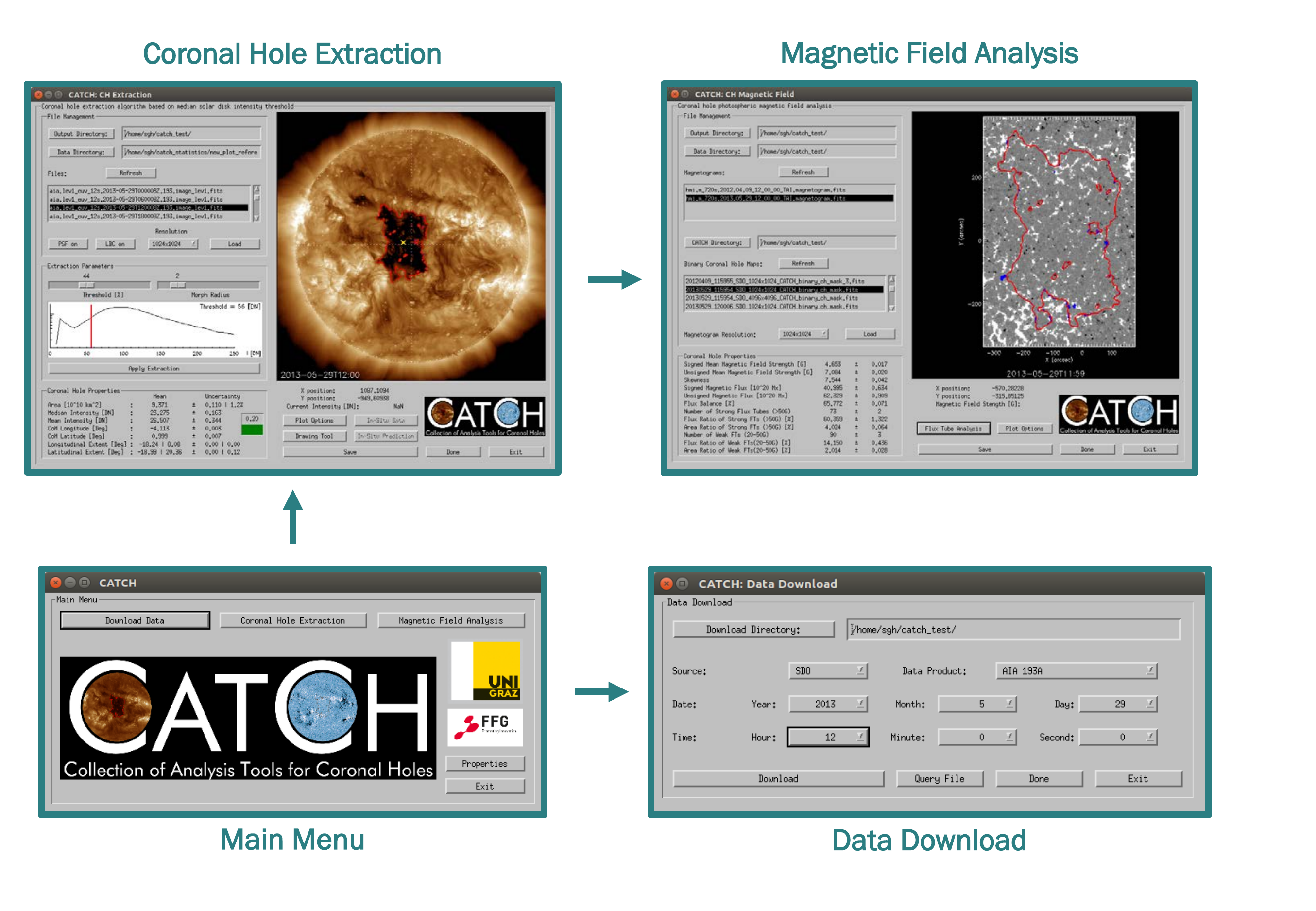}}
 \caption{Screenshots of the CATCH} GUI showing the main menu (left bottom), the data download application (right bottom), the coronal hole extraction option (left top) and the option for the magnetic field analysis (right top).\label{fig:catch}
 \end{figure}

For proper image processing and analysis the SSW (SolarSoftWare) package under IDL (Interactive Data Language) is required, therefore the tool is written in SSW-IDL and the code, including an user-manual, is available on the authors GitHub page (\url{https://github.com/sgheinemann/CATCH}) or by contacting the author directly via E-mail\footnote{For questions, suggestions and the code please contact the main developer S. G. Heinemann via E-mail (stephan.heinemann@hmail.at).}. Figure~\ref{fig:catch} shows the GUI structure of CATCH, displaying the main menu, the data download widget as well as the CH extraction and the magnetic field analysis widget. A more detailed description of CATCH and its functionalities can be found in the user-manual.

\begin{table}
\caption{Parameters calculated with CATCH.}\label{tab:params}
\begin{threeparttable}
\begin{tabular}{l|l|l}

Parameter\tnote{a} &  Unit & Description\\ \hline   
$A_{\mathrm{CH}}$ & km$^{2}$ & Deprojected CH Area\tnote{b} \\
$\bar{I} $& DN &  Mean EUV $193/195$\AA\ Intensity \\
$\widetilde{I} $& DN &  Median EUV $193/195$\AA\ Intensity \\
$\lambda_{\mathrm{CoM}}$ & $^{\circ}$ & Longitude of the Center of Mass (CoM) of the CH           \\
$\lambda_{\mathrm{+}}$ & $^{\circ}$ & Maximum Longitudinal Westward extent \\
$\lambda_{\mathrm{-}}$ & $^{\circ}$ & Maximum Longitudinal Eastward extent \\
$\varphi_{\mathrm{CoM}}$ & $^{\circ}$ & Latitude of the Center of Mass of the CH \\
$\varphi_{\mathrm{+}}$ & $^{\circ}$ & Maximum Latitudinal Northward extent \\
$\varphi_{\mathrm{-}}$ & $^{\circ}$ & Maximum Latitudinal Southward extent \\
$\zeta $&  & Category factor: an estimate of the boundary stability \\
\hline
$\bar{B}$ & G & Signed mean magnetic field strength \\
$\bar{B}_{\mathrm{us}}$ & G & Unsigned mean magnetic field strength \\
$\gamma_{B}$ & & Skewness of the magnetic field distribution \\
$\Phi_{\mathrm{s}}$ & Mx & Signed magnetic flux \\
$\Phi_{\mathrm{us}}$ & Mx & Unsigned magnetic flux \\
$R_{\Phi} $& $\%$ &  Flux balance: ratio of signed to unsigned magnetic flux \\
$N_{\mathrm{FT}} $& Nr &  Flux Tube Number \\
$r_{\Phi} $& $\%$ &  Flux ratio: ratio of signed flux from FTs to the signed CH flux \\
$r_{A} $& $\%$ &  Area ratio: ratio of area of FTs to the CH area \\
\hline
\end{tabular}
\begin{tablenotes}
          \item[a] Note, that all magnetic field parameters are calculated using Line-of-Sight magnetograms, which have been corrected for the assumption of radial magnetic field: $B_{\mathrm{i,corr}}=\frac{B_{i}}{\cos(\alpha_{i})}$.
            \item[b] The deprojection was done using a pixel wise correction with $A_{\mathrm{i,corr}}=\frac{A_{i}}{\cos(\alpha_{i})}$ and $\alpha$ being the angular distance from the disk center.
        \end{tablenotes}
\end{threeparttable}
\end{table}

\subsection{Coronal Hole Extraction from EUV 193/195 \AA\ Filtergrams}\label{subs:euv}

\subsubsection{Intensity Threshold}\label{ssubs:int_thr}
The basic principle under which CH extraction operates is an intensity-based threshold technique applied to EUV filtergrams of sufficient contrast, which was developed by \cite{2012rotter}. To find an optimal threshold \cite{2009krista} derived that an intensity distribution of the solar disk (or a subfield) with a CH present differs significantly from a distribution where CHs are absent. Figure~\ref{fig:euvhist} shows as an example the intensity distribution of the solar disk on May 29, 2013. Hereby, the first maximum, seen at lower intensities, represents one or multiple dark structures on the solar disk. It was proposed that an optimal threshold for a CH boundary lies somewhere in the following minimum. However, note that this characteristic shape is often not well established, especially if no large and well defined CHs are present on the solar disk. Also, it has been found that there is a strong Solar Cycle dependence of the solar disk EUV intensity distribution, which is additionally amended by the current conditions on the Sun (\textit{e.g.,} increased abundance of dark structures or bright active regions). As such, neither a fixed threshold nor a median-intensity dependent threshold, which aims to mitigate intensity variation, perform continuously well. Frequent manual adjustments are needed for optimized results. Thus, the aim is to use an  adjustable threshold depending on the current solar conditions, both locally and globally.

\subsubsection{Intensity Gradient, Uncertainty Estimation and Calculation of CH Properties}\label{ssubs:grad_uncertainty}
\begin{figure} 
 \centerline{\includegraphics[width=1\textwidth,clip=]{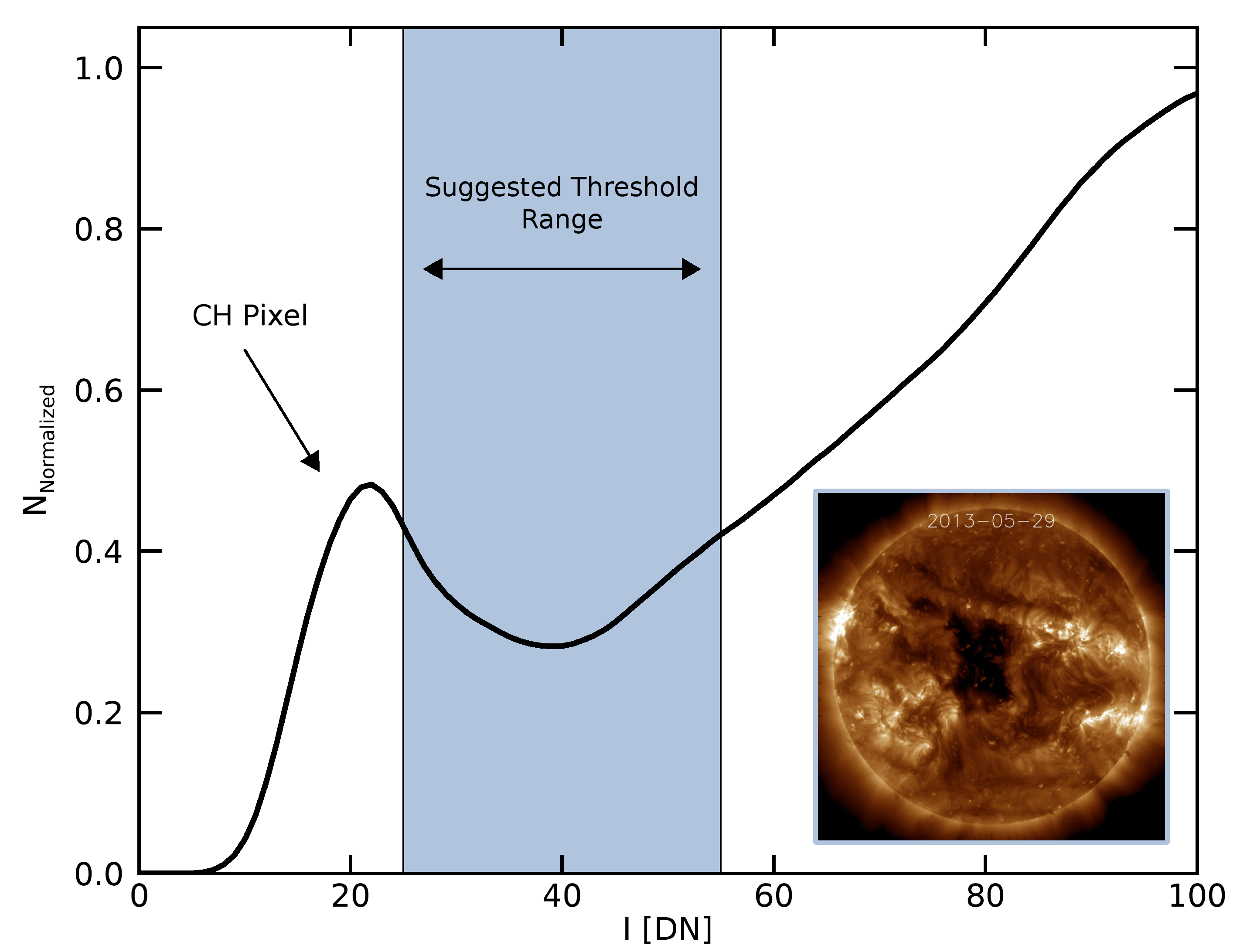}}
 \caption{Normalized distribution of AIA/SDO $193$ \AA\ intensities of the solar disk on May 29, 2013 12UT (see inset). The maximum around $20$ DN represents pixel located inside the CH boundary and the shaded area the reasonable threshold range as proposed by \cite{2009krista}. }\label{fig:euvhist}
 \end{figure}

\begin{figure} 
 \centerline{\includegraphics[width=1\textwidth,clip=]{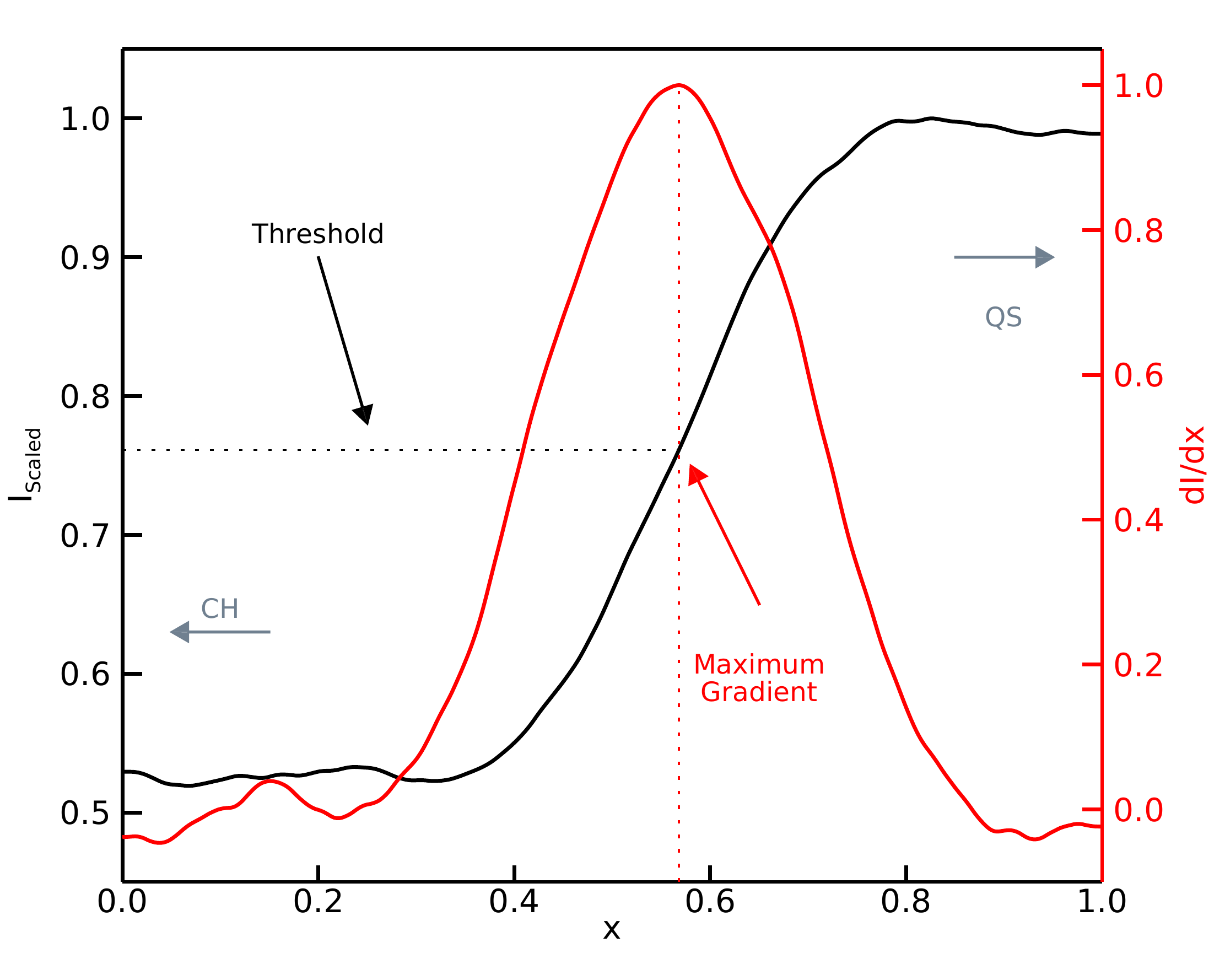}}
 \caption{A representative intensity profile perpendicular to the CH boundary and its derivative. Both are scaled to the maximum ($I_{max}=1$). The $x$-axis represents the radial distance from inside of the CH ($x=0$) across the boundary to the surrounding quiet Sun ($x=1$).}\label{fig:ch_boundary_gradient}
 \end{figure}

The common intensity-based methods have the drawback that the threshold range in which the boundary is considered optimal, is large (see Figure~\ref{fig:euvhist}, shaded area). To narrow down the range of reasonable thresholds, we propose an \textit{intensity gradient} method to estimate the boundary stability and give relevant errors to calculated properties.
Recent studies, investigating CHs and their boundaries, have revealed a steep intensity gradient at the CH boundary \citep{2017hofmeister}. This is due to a strong decrease of the plasma density of quiet Sun temperatures around $1.6$ MK \citep[][]{2011hahn}. Figure~\ref{fig:ch_boundary_gradient} shows a representative intensity profile perpendicular to the CH boundary layer, from inside of the CH ($x=0$) to outside ($x=1$) in arbitrary scale. The y-axis shows the intensity that is scaled to the maximum in this interval which represents the quiet Sun intensity. We see that within a small layer the intensity drops by at least $40\%$ from the quiet Sun level. This small layer represents the range where CH boundaries are usually extracted. Assuming that the CH boundary is best represented where the intensity profile is changing most strongly, we define the optimum boundary to be placed at the steepest intensity gradient (\textit{i.e.}, gradient has a maximum). In an ideal case, the implication of this definition is that the boundary is approximately constant for small threshold variations around the maximum intensity gradient threshold.
This physical 1D principle of the maximum intensity gradient perpendicular to the boundary can be extended to 2D to consider the entire boundary instead of one localized cross--section. This can be done by calculating the change of the CH area for a given intensity threshold by varying the threshold slightly. Using the assumption of a similar intensity gradient along the full boundary, a minimum in the change of the area indicates that, on average, the boundary is located at the maximum gradient, \textit{i.e.} the optimal threshold. 

With this definition of the boundary we aim to minimize the variations in different parameters (first of all the area) to properly estimate the boundary. Practically, this is done by calculating the parameters not only for the boundary defined by the selected threshold but also for boundaries of slightly larger and smaller thresholds. From this set of boundaries, a mean value ($\bar{P}$) and its uncertainty ($\epsilon_{P}$) is calculated. The uncertainty corresponds to the maximum deviation between the determined values and the mean value. 

A reasonable CH boundary can be determined by finding the threshold that minimizes $\epsilon_{A}$ (uncertainty in the CH area) and the CH properties are then given as:

\begin{equation}\label{eq:prop}
P_{\mathrm{CH}}=\bar{P} \pm \epsilon_{P}.
\end{equation}

\subsection{Analysis of the Underlying Photospheric Magnetic Field}\label{subs:mag}
To extract and investigate CHs, it is not sufficient to only use the information extracted from EUV filtergrams as it lacks information about the underlying magnetic structure. The magnetic field configuration is what distinguishes CHs from other dark structures (\textit{e.g.,} filament channels, coronal dimmings) in the solar corona. Studies suggest that it may be possible to differentiate those structures purely from intensity filtergrams \citep{2014reiss} but a clear distinction cannot always be made. A much more precise approach is the definition based on the underlying magnetic field \citep{2015reiss,2018DELOUILLE}.  CHs are defined by their open magnetic field configuration which is reflected in the ratio of the total signed to the unsigned magnetic flux inside the CH and in the skewness of the magnetic field distribution. Filaments and filament channels on the other hand ideally show a symmetric distribution between pixels of positive and negative magnetic flux (closed magnetic structures), as they are located along polarity inversion lines. Thus, analyzing the magnetic field underlying an extracted dark structure reveals its magnetic configuration and enables a clearer classification as CH or filament. 

The calculation of the photospheric magnetic field underlying a CH is often performed by a simple projection of the EUV extracted boundary onto the photospheric magnetogram (line--of--sight or radial). However, it is important to stress that there are several uncertainties in the extraction. First of all, the height difference between coronal imaging in EUV (EUV $193$ \AA : ~$1.01-1.05$ R$_{\astrosun}$) and photospheric magnetic field ($1.00$ R$_{\astrosun}$). Second, the unknown expansion of the magnetic field over the EUV height. Simple projections will have an increased effect on the CH boundary the further it is located away from the center of the solar disk. Another source of uncertainty arises from the noise level, resolution and smoothing of the magnetogram. This can cause non-trivial effects on parameters like the unsigned magnetic flux, flux balance and skewness of the magnetic field. This complicates a comparison of magnetic field properties derived from differently prepared magnetograms. When interpreting such parameters a relative comparison should be preferred rather than relying on absolute values.

\cite{2019hofmeister} showed that the photospheric magnetic field underlying CHs can be divided into 3 categories:  $ \approx 22 \pm 4\%$ of the signed magnetic flux is contributed by a slightly unbalanced background field.  $ \approx 5 \pm 0.1\%$ come from small scale unipolar magnetic elements (flux tubes, FTs) nearly symmetrically distributed over both polarities and which are associated with the super-, meso-, and granular motion of the photosphere. The major contribution, on average $ \approx  69 \pm 8\%$, comes from strong and long-lived FTs which have almost exclusively the dominant polarity of the CH. To map these properties, we calculate the contribution of FTs to various CH parameters. We define two FT categories, strong and weak (with the category weak also covering medium FTs; for more details see  \citealt{2018heinemann_paperII}). FTs are extracted as structures of pixels above a magnetic field strength of $20$ G and the mean magnetic field strength of each structure determines the category. If the mean magnetic field strength of one FT is between an absolute value of $20$ to $50$ G it is categorized as weak, if exceeding $50$ G then it is considered strong.

\section{Statistical Analysis}\label{s:data}
\subsection{Data and Data Processing}\label{ssubs:data}

For the presented statistical study, we did not exhaust all the possibilities of CATCH but constrained the used dataset to one spacecraft. SDO was chosen over STEREO because of the availability of magnetic field maps, and over SOHO because of the better resolution and contrast. The dataset ranges from May 2010 until February 2019. The EUV $193$\AA\ filtergrams observed by AIA/SDO as well LoS magnetograms from HMI/SDO were acquired in a 1 day cadence using the Joint Science Operations Center Servers via the CATCH download application. For the magnetograms the 720s LoS data product was preferred over the 45s due to the lower photon noise of $\approx3$ G measured near the disk center and a better signal-to-noise ratio \citep{2016couvidat_HMI}.

The EUV filtergrams and magnetograms were prepared to level 1.5 using standard SSW-IDL routines and the EUV filtergrams were down-scaled from a pixel scale of $4096 \times 4096$ to $1024 \times 1024$ to significantly enhance the processing speed. Before the extraction, the full--disk filtergrams were corrected for limb-brightening using the anulus limb brightening correction \citep{2014spoca} which is available in CATCH. The boundaries were smoothed using circular ($2$--pixel radius) morphological operators (open and close). To avoid the loss of information on the magnetic fine structure, the magnetograms were not down-scaled. The EUV extracted boundaries were re-scaled to fit the magnetograms resolution.

Note that the effects on boundary detection as well as on the calculation of the parameters in the EUV due to down-scaling are negligible. \textit{E.g.}, we tested for an isolated CH, located close to the disk center on May 29, 2013 how the area of the extracted CH changes for a fixed threshold ($43\%$ of the median solar disk intensity). By varying only the resolutions between $4096 \times 4096$ and $1024 \times 1024$, (without smoothing) we find a deviation of the extracted CH of less than $0.5\%$. Other parameter behave similar. As such, the uncertainties from using different filtergram resolutions to extract CH boundaries is much lower than uncertainties in the extraction itself.

From the daily EUV images, dark structures located close to the central meridian were extracted (Center of Mass, CoM located $\pm 10^{o}$). The extracted structures were limited to the central meridian to reduce longitudinal projection effects due to the spherical nature of the Sun. Polar CHs as well as polar connected CHs were excluded for the same reason. Each structure was extracted only once for each solar disk passage to avoid statistical biases because of similar datapoints. The magnetic properties of each dark structure were investigated and non-CH structures were identified (defined as structures with a flux balance below $10\%$ or a magnetic field skewness below $1$) and discarded from further analysis. This approach yielded $707$ CHs over a wide range of sizes and latitudes spanning a timerange of more than $8$ years.

\subsection{Results}\label{subs:res}

We analyzed $707$ CHs near their central meridian passage and categorized them by their boundary stability. All the parameters presented here, are calculated with CATCH. Our findings are as follows:

\subsubsection{Assessment of the Stability of the Extracted Boundaries}
\begin{figure} 
 \centerline{\includegraphics[width=1\textwidth,clip=]{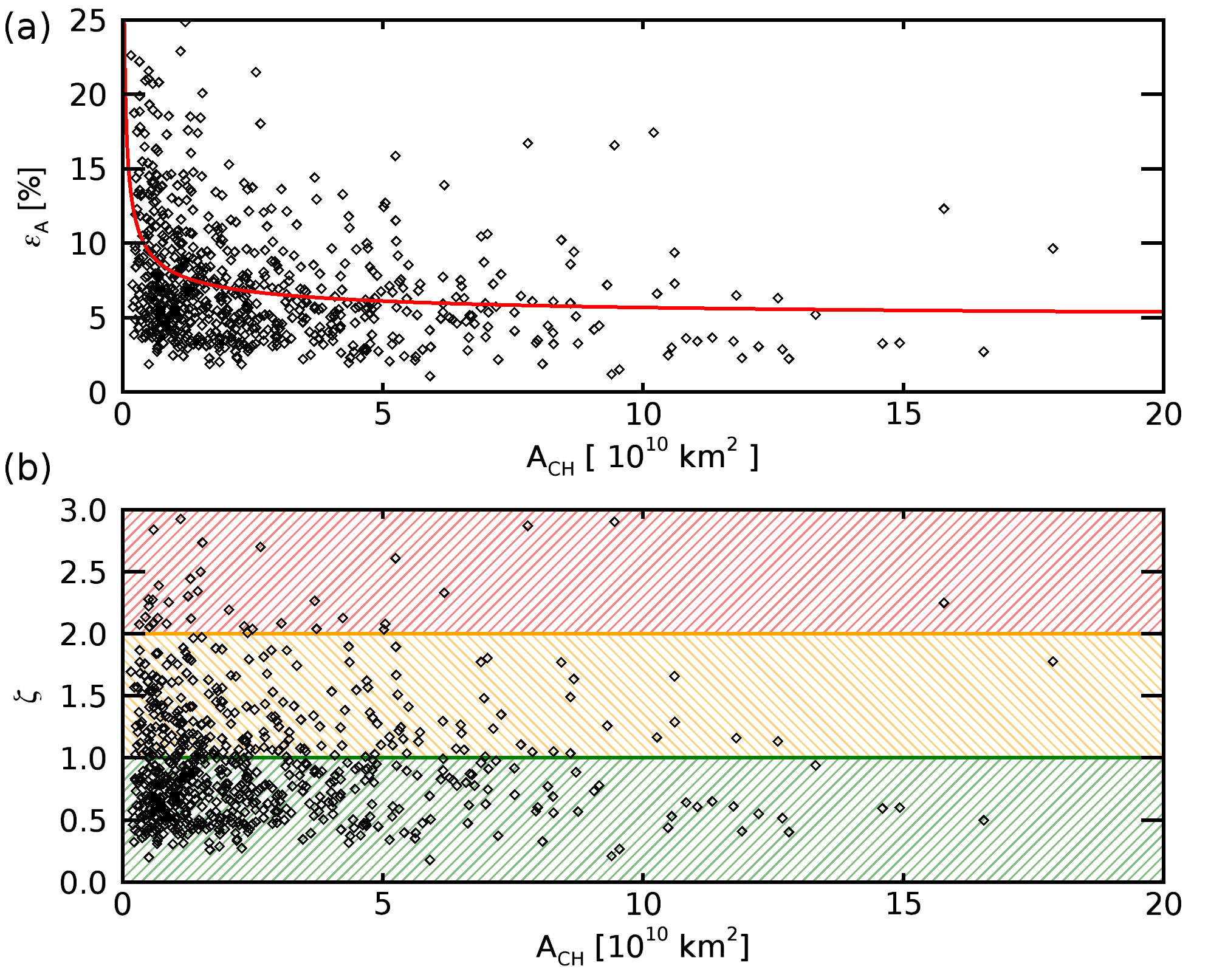}}
 \caption{Panel (a) shows a scatterplot of the CH area against its maximum deviation as given in Section~\ref{subs:euv}. The red line shows the fit which is used to calculate the category factor ($\zeta$). $\zeta$ is plotted against the area in panel (b). The shaded areas (green, orange and red) represent the stability assessment of the boundaries as high, medium and low respectively.}\label{fig:error}
 \end{figure}

First, we assessed the stability of the extracted CH boundaries by analyzing $\epsilon_{A}$ for the optimal threshold for all $707$ CHs. Figure~\ref{fig:error}a shows the CH area ($A_{\mathrm{CH}}$) against its uncertainty ($\epsilon_{A}$). We find a dependence on the area which seems to have two causes: (1) the larger impact of stray light for smaller CHs which could partly  be compensated by performing a PSF deconvolution before the CH extraction and (2) the non-zero extent of the boundary layer whose area is growing linearly in contrast to the total CH area (which grows according to a power law). This causes larger percentage variation for smaller CH areas. To correct for this dependence we introduce the category factor ($\zeta$) which can be given as:
\begin{equation}\label{eq:zeta}
    \zeta=\frac{\epsilon_{A}}{f_{\mathrm{fit}}(A_{\mathrm{CH}})},
\end{equation}
with $f_{\mathrm{fit}}(A)$ being the fit shown in Figure~\ref{fig:error}a as the red line. It is given by:
\begin{equation}
   f_{\mathrm{fit}}=3.31\times (A_{\mathrm{CH}})^{-0.53}+4.71,
\end{equation}
with $A_{\mathrm{CH}}$ in units of $10^{10}$ km$^{2}$. The resulting $\zeta$-factor as function of CH area is shown in Figure~\ref{fig:error}b. From this we define three categories of boundary stability:
\begin{enumerate}
    \item high: $\zeta \le 1$
    \item medium: $1 < \zeta \le 2$
    \item low: $\zeta > 2$
\end{enumerate}
 We find that $60.0\%$ of the CHs under study have a high boundary stability, $34.2\%$ a medium and only $5.8\%$ are of low boundary stability.

\subsubsection{Thresholds}\label{ssubs:thr}
\begin{figure} 
 \centerline{\includegraphics[width=1\textwidth,clip=]{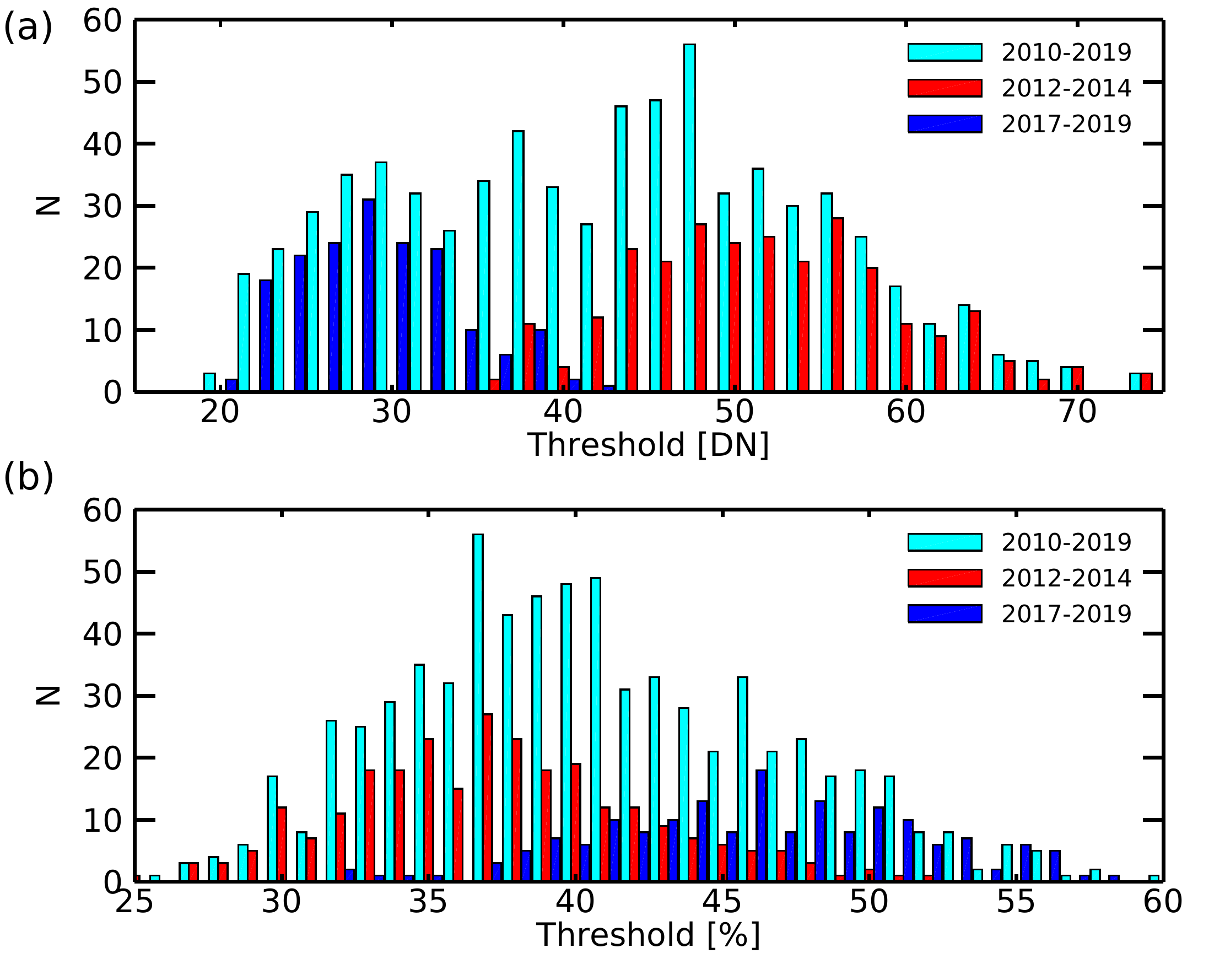}}
 \caption{Distribution of the optimal threshold during the entire SDO-era (cyan), covering solar maximum (red) and the decline and minimum phase (blue). Panel (a) shows the threshold in absolute counts (DN) and panel (b) in percent of the median solar disk intensity.}\label{fig:thr_distr}
 \end{figure}

\begin{figure} 
 \centerline{\includegraphics[width=1\textwidth,clip=]{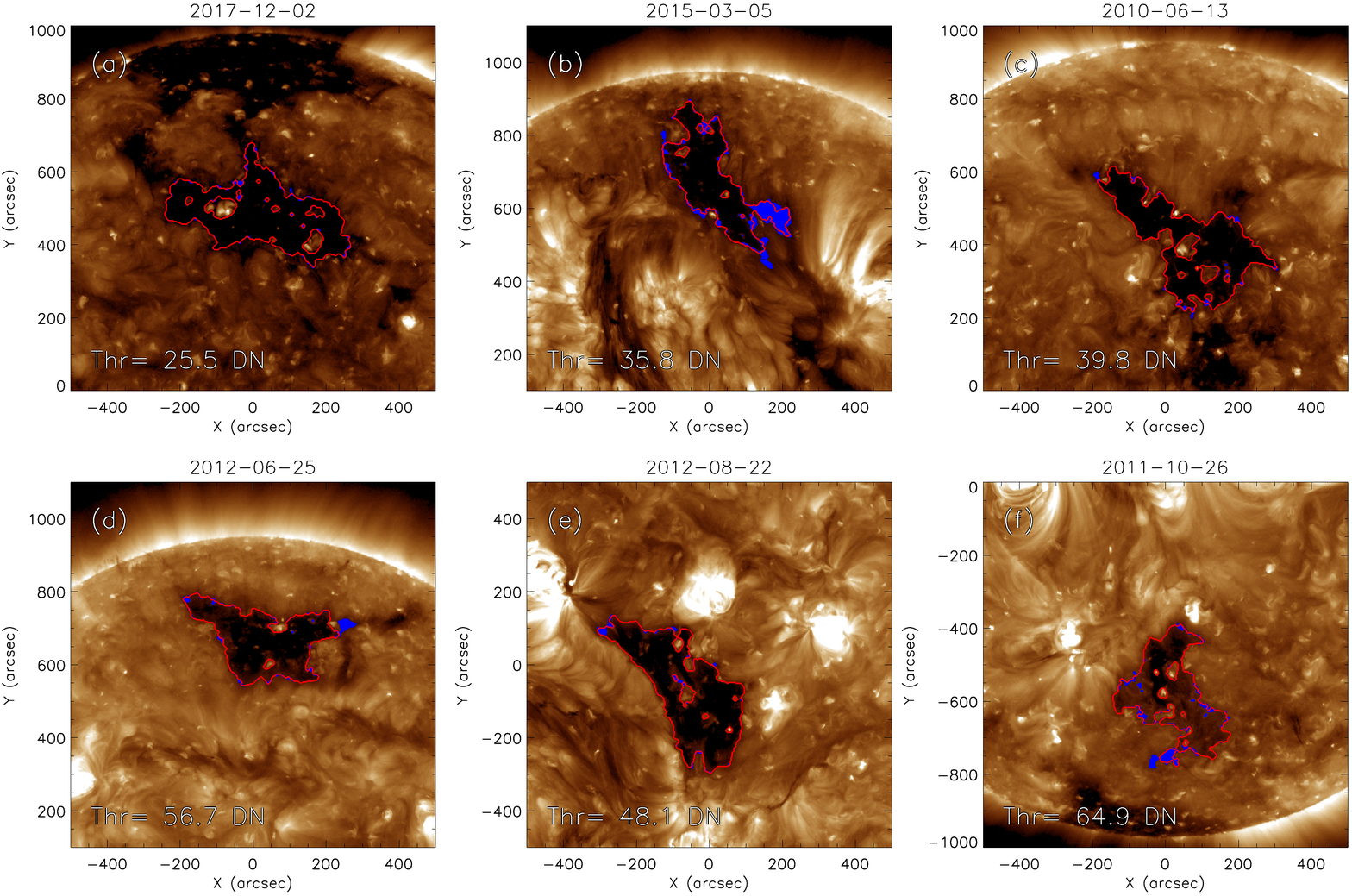}}
 \caption{Sample of CHs extracted with varying optimal thresholds (as defined in Section~\ref{subs:euv}). The sample represents the threshold value distribution as given in Figure~\ref{fig:thr_distr}. The optimal thresholds range form $25$ to $65$ DN and are primarily caused by a different intensity quiet Sun level rather than large changes in the CH intensity. The red boundary corresponds to the boundary derived by the optimal threshold and the blue shaded areas are the uncertainties (see Section~\ref{ssubs:grad_uncertainty}). All images are equally scaled.} \label{fig:thr-fig}
 \end{figure}

Second, we investigated how the optimal threshold to extract CHs is distributed and varies over the course of the observed time period from 2010 to 2019. This period nearly covers the whole Solar Cycle 24. Figure~\ref{fig:thr_evo} shows the threshold over time (a) in absolute counts (DN) and (b) in percent of the median intensity of the solar disk. The black line in panel (d) shows the smoothed daily sunspot number by SIDC/SILSO\footnote{The daily sunspot number can be found via SIDC/SILSO \url{http://www.sidc.be/silso/}.}, which acts as a proxy of the solar activity. We find a clear Solar Cycle dependence in the optimal threshold (between $25-55$ DN) which cannot be correctly mitigated by modulation with the median solar disk intensity (of the full solar disk). It seems that the correction is too strong, especially during solar minimum. Additionally, because of the individual configuration of CHs, the optimal threshold may vary by up to $\approx 20$ DN for any given time. The distribution of thresholds (in DN) is shown in Figure~\ref{fig:thr_distr}a (cyan) with a mean of $43.9$ DN and a standard deviation of $12.1$ DN. The distribution shifts from $53.5 \pm 8.6$ DN during solar maximum (red, $2012 - 2014$) to $29.7 \pm 4.6$ DN during decline and minimum (blue, 2017-2019). When considering the threshold in percent of the median solar disk intensity (Figure~\ref{fig:thr_distr}b), the mean threshold is $40.4 \pm 6.3 \%$ with a variation between the solar maximum (red, $2012 - 2014$) with $37.3 \pm 5.0 \%$ and the decline and minimum (blue, 2017-2019) with $46.2 \pm 5.5 \%$. We find the threshold to be independent of the CH size. We believe that the large change of the optimal threshold (in DN) is due to the change in intensity due to the Solar Cycle evolution (\textit{e.g.,} number of active regions, higher quiet Sun level, ...). Figure~\ref{fig:thr-fig} shows six CHs extracted with a different optimal threshold, varying from $25$ to $65$ DN.

\subsubsection{Area, Intensity, and Position}\label{ssubs:a,i,p}
\begin{figure} 
 \centerline{\includegraphics[width=1\textwidth,clip=]{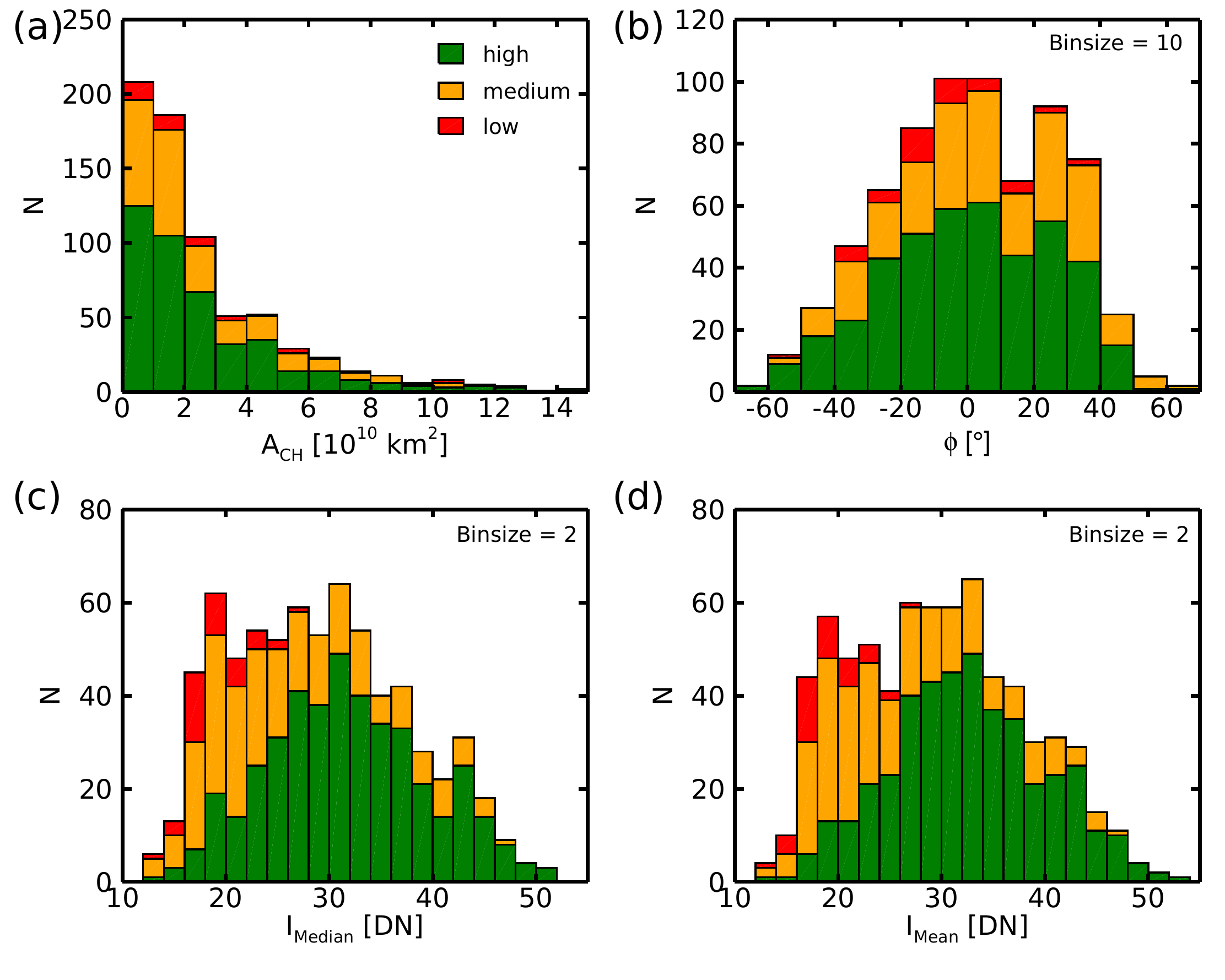}}
 \caption{Distribution of CH properties sorted corresponding to the category factor (green: high, orange: medium and red: low). Panel (a) show the area distribution, panel (b) the distribution of the latitudinal location of the center of mass and the panels (c) and (d) show the distribution of the median and mean intensity within the extracted CH boundaries.}\label{fig:hist_prop}
 \end{figure}

After investigating the extraction mechanism in terms of intensity threshold, we analyze how CH properties are distributed in our data-set.  Figure~\ref{fig:hist_prop} gives an overview of the main CH properties, \textit{i.e.}, the distribution of the areas, latitudes, and intensities of all CHs under study. Figure~\ref{fig:hist_prop}a shows the distribution of the deprojected areas. We find CH areas ranging from $1.6 \times10^{9}$ km$^{2}$ to $1.8 \times10^{11}$ km$^{2}$, with an average of $(2.69 \pm 2.73) \times10^{10}$ km$^{2}$. CHs with an area below $2\times10^{10}$ km$^{2}$ account for $56\%$ of all CHs, whereas only $5\%$ of CHs exceed an area of $8\times10^{10}$ km$^{2}$. 

The CoM of CHs under study are distributed over latitudes ranging from $-63^{\mathrm{o}}$ to $+63^{\mathrm{o}}$ (Figure~\ref{fig:hist_prop}b). $39\%$ of all CHs, which are located between an absolute value of $40^{o}$ and $20^{o}$, can be considered medium-latitude CHs and $50\%$ are considered low-latitude CHs, located below $20^{o}$. We find the CHs to be nearly balanced between the hemispheres (South: $48\%$ CHs; North: $52\%$ CHs) without a clear relation to the solar activity (see Figure~\ref{fig:thr_evo}c).

We calculate the median and mean intensity in the $193$ \AA\ wavelength for each CH of the dataset. The mean of the median intensities is calculated to be $29.0 \pm 8.5 $ DN (Figure~\ref{fig:hist_prop}c) and the mean of the mean intensities is $29.5 \pm 8.4 $ DN (Figure~\ref{fig:hist_prop}d). When only considering the $50\%$ and $25\%$ pixel with the lowest intensities we find the mean intensity to be $ 23.9 \pm 7.4 $ DN and $ 21.2 \pm 6.9 $ DN respectively.

\begin{figure} 
 \centerline{\includegraphics[width=1\textwidth,clip=]{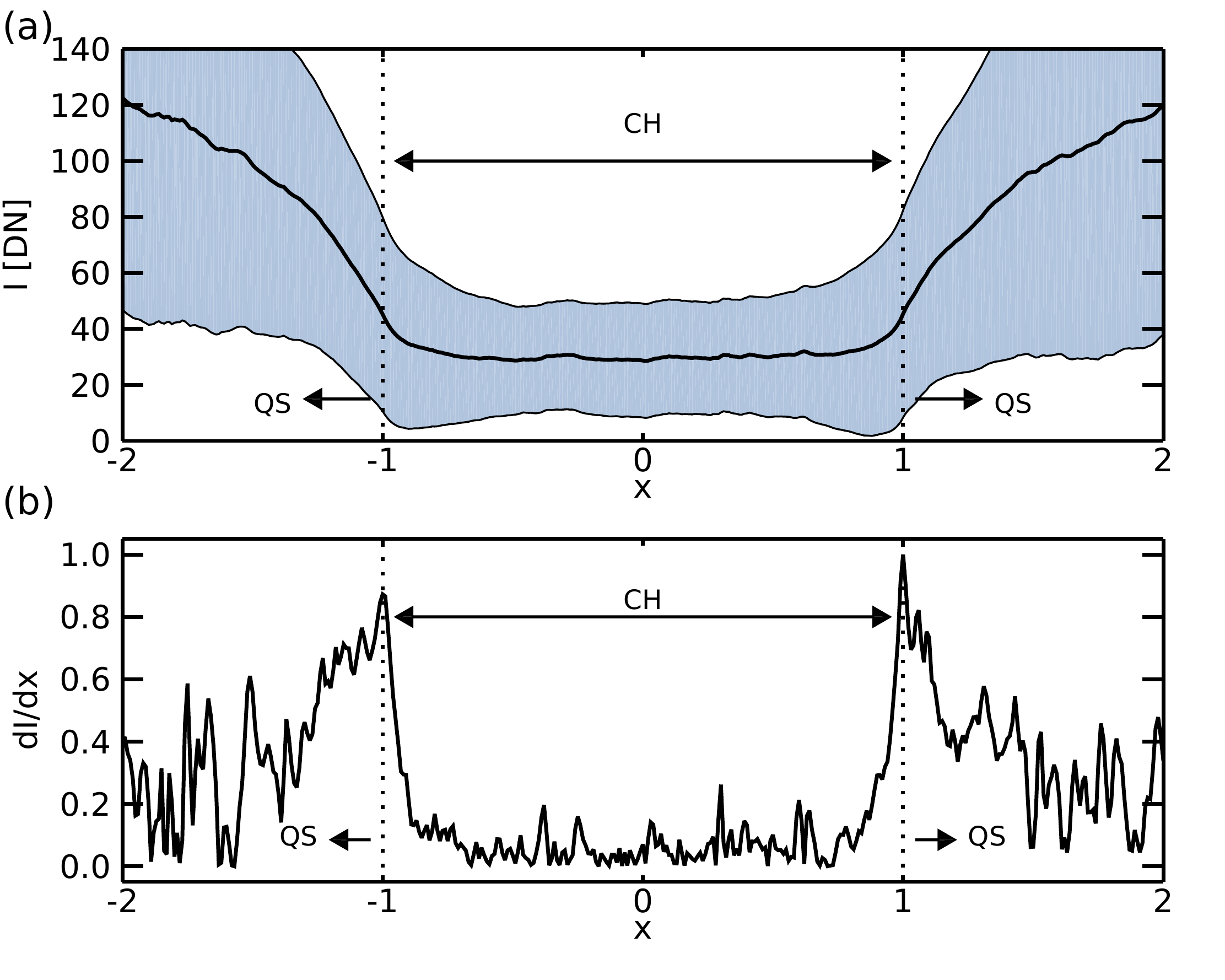}}
 \caption{The superposed intensity profile of the longitudinal cross-sections at the CoM of the CHs under study (a) and its derivative (b). Before superposing, each intensity profile is scaled so that $x\pm1$ represents the CH boundaries. The black line is the mean profile and the shaded gray-blue area represents the $1\sigma$ standard deviation. The dotted vertical lines mark the location of the CH boundary. }\label{fig:crosssection}
 \end{figure}

We investigated the intensity profile of the cross-section of the CHs. To that end we cut the CHs longitudinally through their CoM and superpose the intensity profiles. Figure~\ref{fig:crosssection}a shows the superposed mean profile (black line) with the $1\sigma$ uncertainties represented by the shaded area and the second panel (b) shows the derivative of the mean profile. Note, that the intensity profiles were scaled so that the CH boundaries correspond to $x=\pm1$. We find that when using CATCH to extract CHs the boundary is consequently extracted at the highest gradient in the intensity, which was the initial assumption. With this we can highlight the CH extraction according to a physical principle in contrast to an arbitrarily chosen (or empirically found) value. 

\subsubsection{Properties of the Underlying Photospheric Magnetic Field}\label{ssubs:magres}
\begin{figure} 
 \centerline{\includegraphics[width=1\textwidth,clip=]{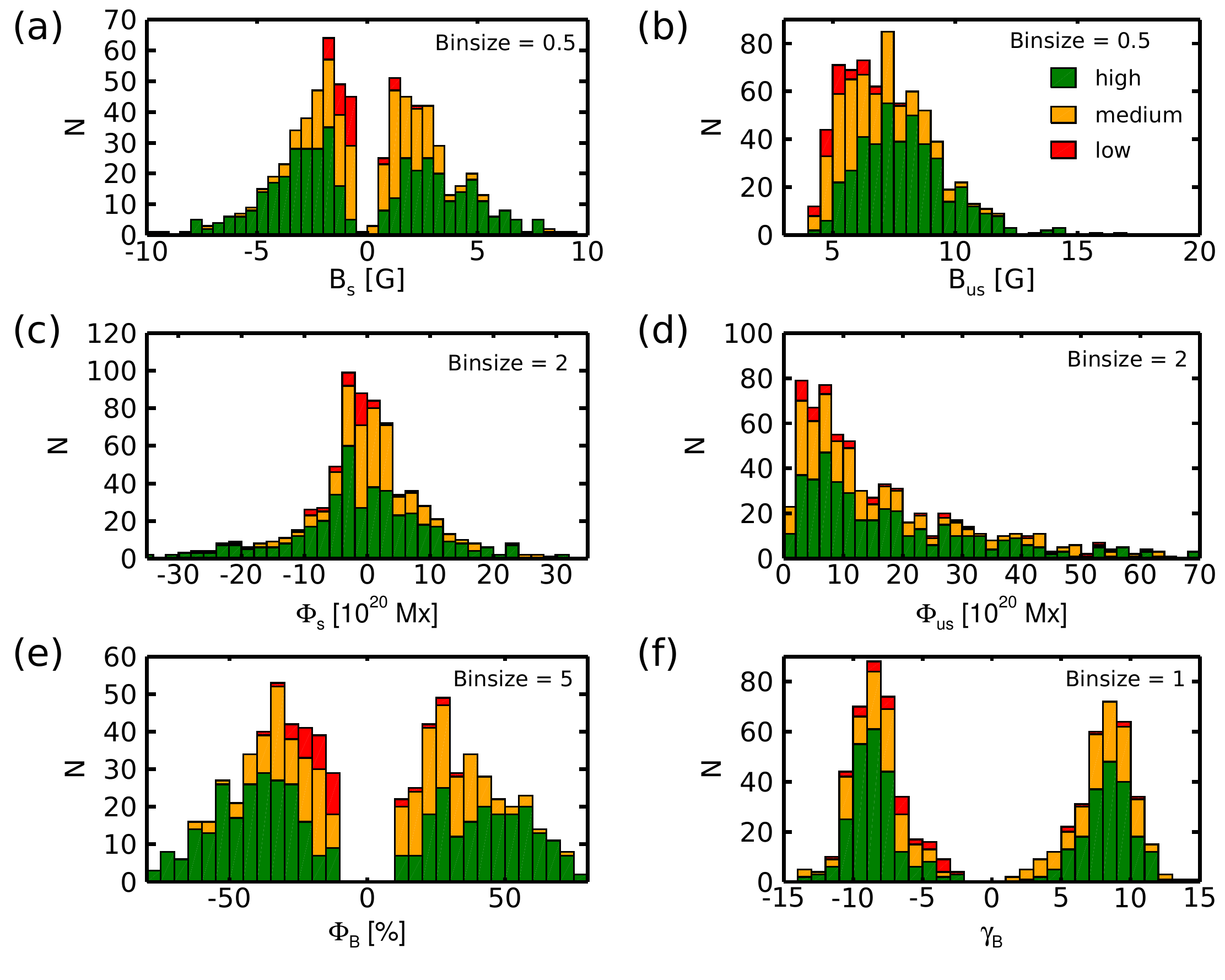}}
 \caption{Distribution of magnetic CH properties sorted corresponding to the category factor (green: high, orange: medium and red: low). Panels (a) and (b) show the distribution of the signed and unsigned mean magnetic field strength of the photospheric field below the CH. The distributions of the signed and unsigned flux are shown in the panels (c) and (d). The flux balance, the ratio between the signed and unsigned magnetic flux is shown in panel (e). Panel (f) shows the distribution of the values for the skewness of the magnetic field distribution.}\label{fig:hist_mag}
 \end{figure}

\begin{figure} 
 \centerline{\includegraphics[width=1\textwidth,clip=]{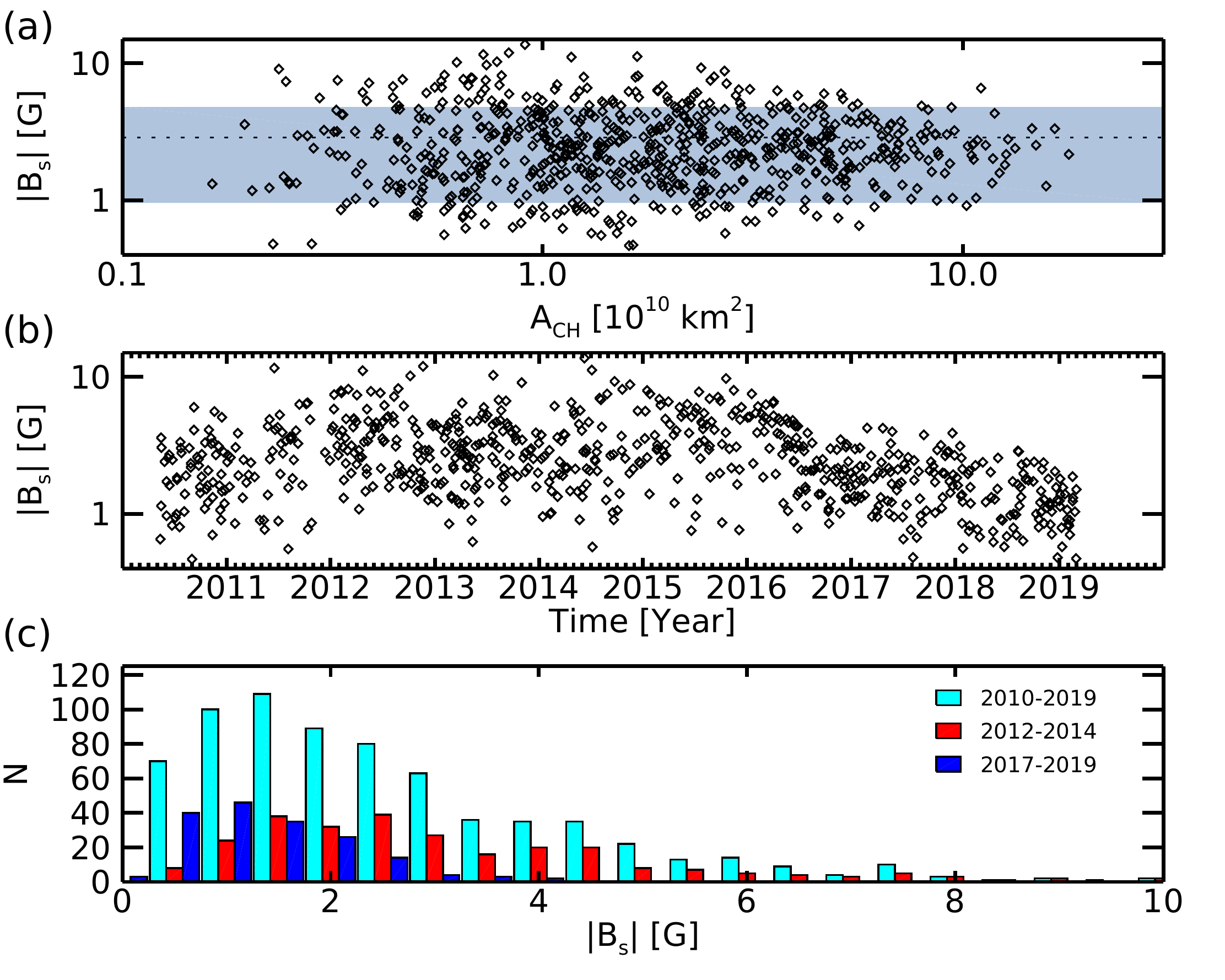}}
 \caption{In panel (a) the CH area is plotted against the absolute value of the signed mean magnetic field strength ($|B_{s}|$) in double logarithmic depiction. Panel (b) gives the temporal evolution of the absolute value of the signed mean magnetic field strength (y-axis is logarithmically scaled). Panel (c) shows the distribution of $|B_{s}|$ for the whole dataset in cyan, for the solar maximum in red and for the declining phase and the minimum in blue. }\label{fig:mag_evo}
 \end{figure}

The analysis of the magnetic field properties underlying the CHs yielded a near symmetric distribution of positive and negative polarity CHs (Figure~\ref{fig:hist_mag}a). The mean of the absolute values of the signed mean magnetic field strength ($|B_{s}|$) is $2.9\pm 1.9$ G (Figure~\ref{fig:mag_evo}c). There seems to be no correlation between $|B_{s}|$ and the CH area (Figure~\ref{fig:mag_evo}a; see also Figure~\ref{fig:corrmatrix}). There is however a slight difference in the distribution of $|B_{s}|$ between the solar maximum against the decaying and minimum phase. In the maximum ($2012 - 2014$) the absolute value of the mean magnetic field strength exhibits a slightly higher average and a wider spread with $3.4 \pm 2.1$ G than in the decaying phase and solar minimum (2017-2019) with $1.6 \pm 0.8$ G (Figure~\ref{fig:mag_evo}b,c). The increased magnetic field strengths within CHs during solar maximum may be the result of enhanced magnetic activity during the reversal of the solar magnetic field which causes more active regions to appear and consequently decay \citep{2010Karachik}. In Figure~\ref{fig:hist_mag}b the unsigned mean magnetic field strength is shown. We find the mean to be $7.3 \pm 1.9 $ G and that $90\%$ of the CHs have a value below $10$ G. We note, that the unsigned magnetic flux is strongly dependent on the magnetogram resolution and smoothing, as it is dominated by the sum of the noise of the magnetic field pixel distribution. Therefore, the unsigned magnetic flux should be considered in relative comparison (\textit{e.g.,} between two CHs) rather than in absolute values. This is also true for the skewness and the flux balance. \textit{E.g.}, for the isolated CH located near the disk center on May 29, 2013 the skewness changes from $9.2$ at a resolution of $4096 \times 4096$ to $7.4$ at a resolution of $1024 \times 1024$, and the flux balance changes from $57.2\%$ to $66.7\%$, respectively. 

The signed magnetic fluxes of the CHs seem to be symmetrically distributed between both polarities. The mean of the absolute value is given at $(7.2 \pm 13.5) \times 10^{20}$ Mx with a maximum value of $6.9 \times 10^{21}$ Mx. The unsigned magnetic fluxes range from $7.2 \times 10^{19} $ to $ 2.0 \times 10^{22} $ Mx with a mean of $(2.0 \pm 3.7) \times 10^{21}$ Mx (Figure~\ref{fig:hist_mag}c,d). The flux balance, the ratio of the signed magnetic flux to the unsigned magnetic flux and with that hinting towards open magnetic flux, shows that the CHs are distributed from $10\%$ to $87\%$ with a mean of $36,3 \pm 16.3\%$ (Figure~\ref{fig:hist_mag}e). The $47\%$ of CHs that show positive polarity have a mean flux balance of $37.4\pm 16.7 \%$ whereas the $53\%$ of CHs that show negative polarity have a mean of $-(35.6 \pm 15.7) \%$. The shift in the magnetic field distribution that characterizes CHs is shown in Figure~\ref{fig:hist_mag}f. The mean of the absolute value of the skewness is $8.2 \pm 2.1 $, clearly showing the asymmetry in the magnetic field caused by the abundance of open magnetic field lines. There is no clear difference between polarities or boundary categories.

\subsubsection{Flux Tubes}\label{subs:ft}
\begin{figure} 
 \centerline{\includegraphics[width=1\textwidth,clip=]{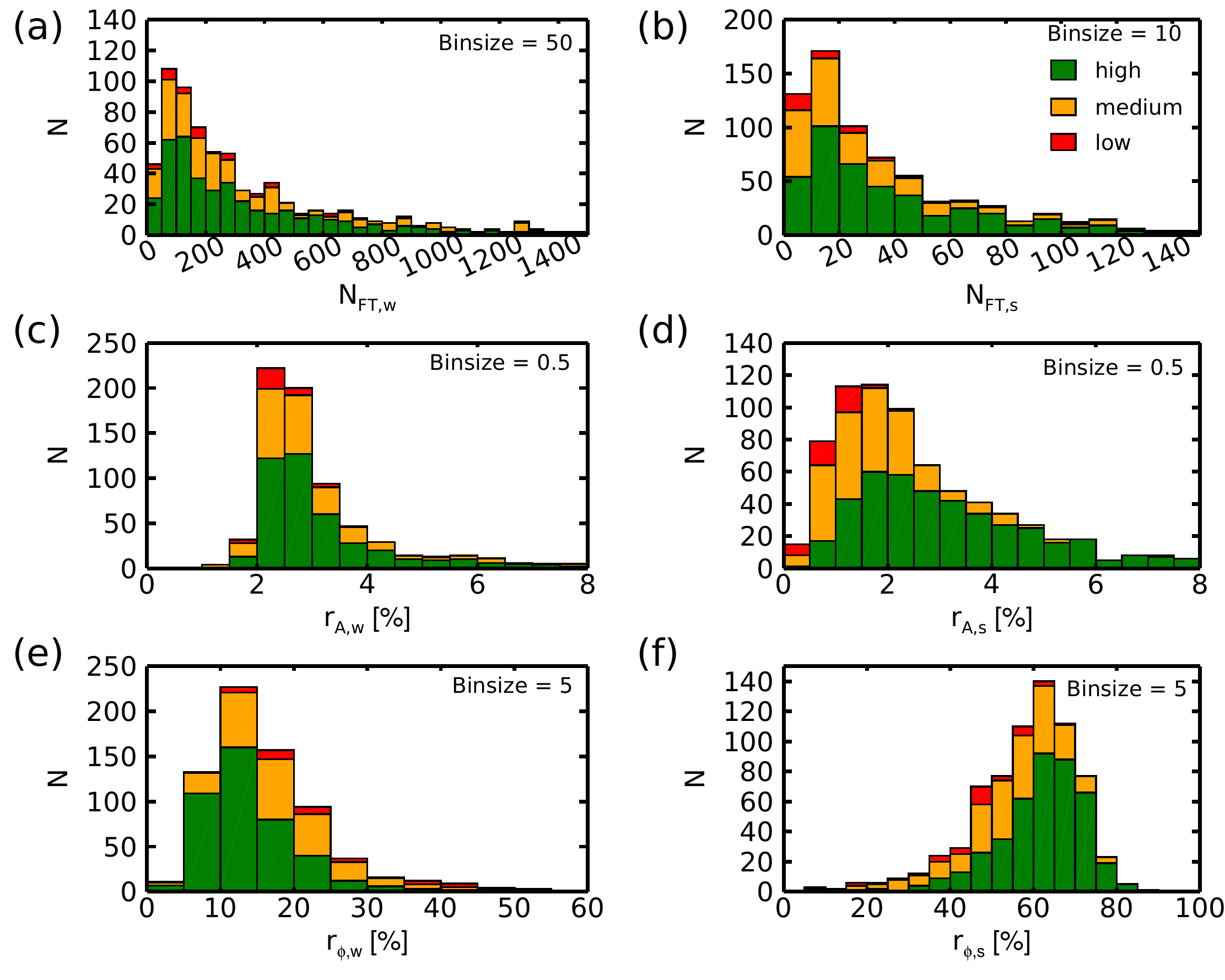}}
 \caption{Distribution of FT properties within the CH sorted corresponding to the category factor (green: high, orange: medium and red: low). Panels (a) and (b) show the distribution of the number of FTs per CH (weak and strong respectively). The distribution of the area ratio of FTs for weak and strong FTs is shown in panel (c) and (d). Panels (e) and (f) show the distribution of the FT flux ratio for weak and strong FTs.}\label{fig:hist_ft}
 \end{figure}

Besides the magnetic parameters for the global structure of a CH, CATCH can analyze the fine structure of the magnetic field in form of FTs or magnetic elements. We analyzed the contribution of the small unipolar FTs categorized as weak ($20$ G $< |B_{s,\mathrm{FT}}| < 50$ G) and strong ($|B_{s,\mathrm{FT}}| > 50$ G) to the CH parameters. Figure~\ref{fig:hist_ft} shows the distribution of FT number (a,b), area ratio (c,d) and flux ratio (e,f) for weak and strong FTs respectively. There is no significant difference for the three categories of boundary stability. The number of extracted FTs ranges from $15$ to $2670$ for the weak and from $1$ to $223$ for the strong FTs. This gives an average of $ 135 \pm 30 $ FTs per $10^{10}$ km$^{2}$ for weak and $15 \pm 6 $ FTs per $10^{10}$ km$^{2}$ for strong FTs. The number of strong FTs per area scales with the signed mean magnetic field strength of the CH (Pearson Correlation Coefficient (cc$_{p}$): $0.74$ with a $95\%$ confidence interval (CI) of $[0.70,0.78]$), whereas the weak ones do not (cc$_{p}=-0.12$ with a $95\%$ CI of $[-0.18,-0.06]$).

When analyzing the contribution of the weak and the strong FTs to the area and signed magnetic flux of the CH we find that the strong ones are dominating. For most CHs ($90\%$) the contribution of the strong FTs to the signed magnetic flux is between $40\%$ and $80\%$ with a mean of $58.1\pm 13.1\%$, although they only cover between $0.5\%$ and $6\%$ of the CHs area (on average $2.6 \pm 1.8\%$). We find that the coverage of the strong FTs is strongly correlated with the mean magnetic field strength of the CH (cc$_{p}=0.98$ with a $95\%$ CI of $[0.97,0.98]$). In contrast, the weak FTs only contribute $16.3 \pm 8.8 \%$ of the signed magnetic flux and cover a rather constant CH area of $1.5-4\%$ ($84\%$ of CHs) without a correlation to the mean magnetic field strength of the CH (cc$_{p}=-0.02$ with a $95\%$ CI of $[-0.07,0.03]$).

\vspace{5mm}

We note that in the Appendix, Figure~\ref{fig:corrmatrix} the pairwise Spearman correlation coefficients of all parameters calculated in the statistical study using CATCH can be viewed. We note, that not all correlations imply a causal relationship, but might be correlated by definition.
 
\section{Discussion}\label{s:disc}
Using the intensity profile perpendicular to the boundary layer of CHs we were able to improve the intensity-based threshold method by \cite{2012rotter}, based on the concept initially proposed by \cite{2009krista}. By adding an estimation of the boundary stability and uncertainty, local as well as global influences on the CH intensity can be described and compensated.  By investigating the performance of the newly adjusted threshold method we highlight the advantages of such a supervised method.

\subsection{CATCH}
Reliable extraction of CHs from EUV observations is an important step towards understanding their configuration, a necessary aspect in solar- and space weather research and space weather applications. Without a precise definition of the CH boundaries, which is then applied to all CHs under study, an analysis is often biased by differences in the extracted boundary and by local conditions which lead to significant problems in the comparison of different studies. Approaches that aim to optimize a threshold for full--disk images \citep{2012rotter,2017hofmeister,2018garton} or synoptic maps \citep{2018hamada} can adjust for global changes in the intensity distribution, but do not take into account the local variations. In another study, a dual--threshold--based approach (\textit{ezseg}: \cite{2016caplan_LBC_IIT})  was developed with the aim to consider local variations, if the threshold pair is properly tuned.



Our analysis yielded that the optimal threshold (as defined in Section~\ref{subs:catch}) between CHs in one single filtergram may vary significantly due to the abundance and proximity of active regions, quiet Sun areas and bright loops. \cite{2018Wendeln} found from differential emission measure analysis (DEM), that a significant contribution within CHs comes from stray light of nearby active regions, high overlying loops and the instrumental point spread function (PSF). It is reasonable to suggest that these effects also influence the CH extraction in one (or multiple) wavelengths. However, by individually assessing the boundary of each CH, effects coming from local conditions can be mitigated. We also tested the influence of the PSF by deconvoluting the images before extraction (this option is available in CATCH using SSW-routines) and found clear enhancement in the extraction process but dismissed it for the statistical study because of the greatly increased processing time (up to a factor $100$). There are faster options to perform PSF deconvolution when not using IDL \citep[\textit{e.g.,}][]{2012Prato}, which have not been explored yet as CATCH is entirely written in SSW--IDL.

The intensity profile perpendicular to the CH boundary is very dependent on the coronal configuration outside the CH. Active regions have significantly higher intensities than the quiet Sun, but also loops associated with activity near the CH boundary show increased intensities. Enhancements near and at the boundaries may be the result of the CH evolution through the process of interchange reconnection \citep{2009madjarska,2010edmondson,2011yang}. It is a known drawback of this method that by considering the average gradient along the full CH boundary, small scale conditions are neglected. The method approximates that the intensity gradient across the boundary of a given CH is constant along the boundary, which we know is not always true. This leads to uncertainties, which we approximate as presented in Section~\ref{subs:euv}. To consider such small scale variations, a much more precise definition of the boundary needs to be established which requires a new approach for detecting CH boundaries. Automated threshold techniques are fast, but may extract several CHs in close vicinity which may or may not appear with merged boundaries depending on slight variations in the threshold. However, the threshold--based method described here delivers consistent results when manually supervised. This is due to the constrains set by the boundary gradient approach. 

\begin{figure} 
 \centerline{\includegraphics[width=1\textwidth,clip=]{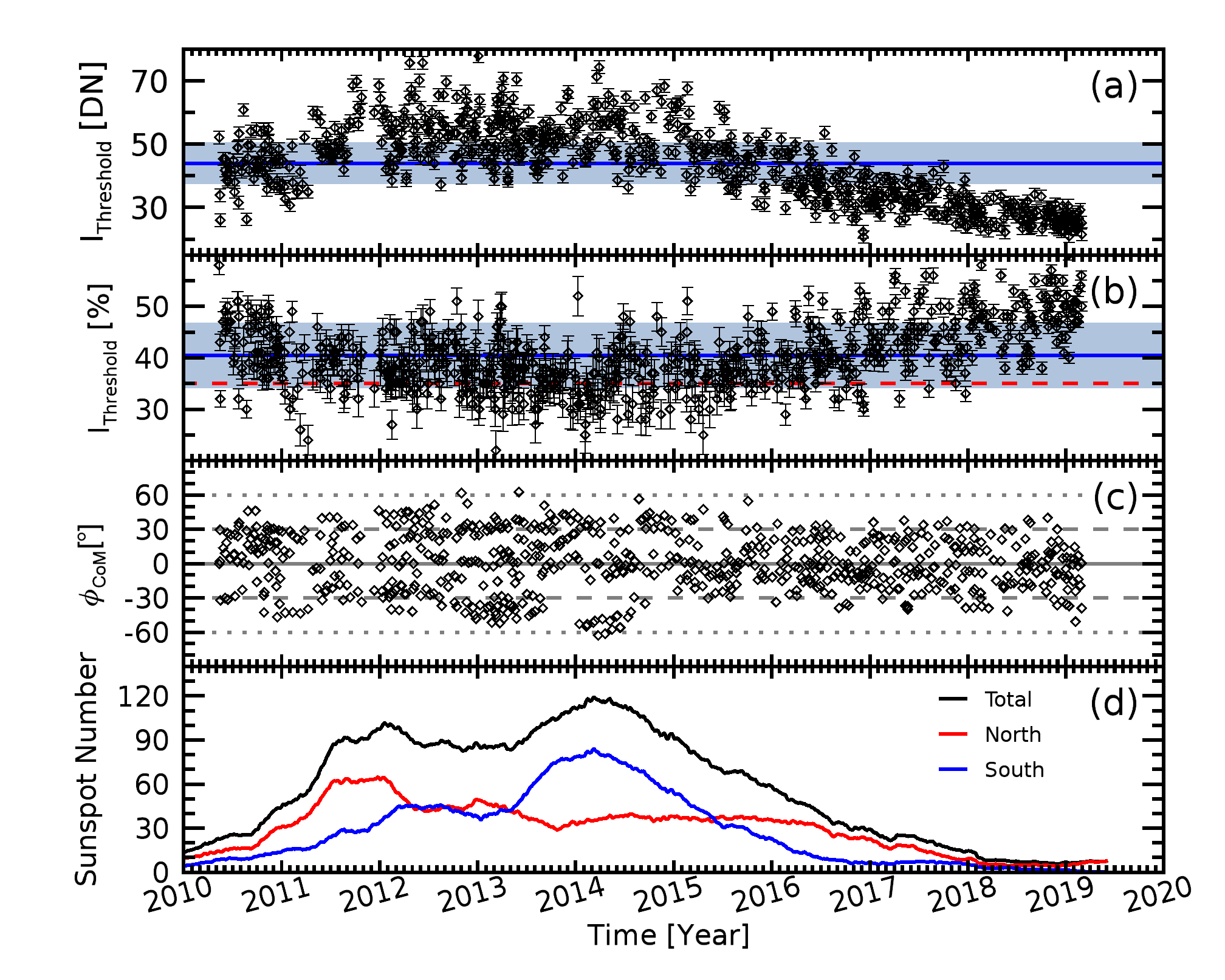}}
 \caption{Evolution of the optimal intensity threshold for CH extraction as function of time. Panel (a) shows the intensity in counts (DN) and panel (b) shows the threshold in percent of the median intensity of the solar disk at the time of the extraction. The blue line represents the mean value and the blue shaded area the $1\sigma$ range over the whole period studied. The dashed-red line in panel (b) represents} the often used threshold value of $35\%$ of the median solar disk intensity. The errorbars indicate the variation of the threshold for the uncertainty estimation as described in Section~\ref{subs:euv}. The third panel (c) shows the latitudinal positions of the CoMs of the CHs under study. The bottom panel (d) shows the smoothed daily sunspot number as provided by the SIDC/SILSO. The black line is the total sunspot number while the red and blue lines show the sunspot number for the northern and southern hemisphere respectively.\label{fig:thr_evo}
 \end{figure}

We therefore pursued to further develop this approach, having in mind the advantages of being computationally very inexpensive, fast and flexible. From the statistical results we derive, we find that our method consistently performs well (by standards of visual inspection) over the changing conditions of a full Solar Cycle and also mitigates local variations. Comparing to the method using a fixed intensity threshold of $35\%$ of the median solar disk intensity  \citep{2007vrsnak,2012rotter,2016reiss,2017hofmeister,2018heinemann_paperI}, we find significant deviations for the boundary we would consider as optimal. This is expected as automated methods are often judged by how close they come to manual or manually--tuned methods.  We also find that the $35\%$ of the median solar disk intensity is a good estimate for the mean threshold during solar maximum (Figure~\ref{fig:thr_evo}b red line; Figure~\ref{fig:thr_distr}). In our study the mean threshold for the time period of the solar maximum ($2012 - 2014$) comes to $37.3 \pm 5.0\%$ of the median solar disk intensity. This is very well shown in the Solar Cycle dependence of the threshold (Figure~\ref{fig:thr_evo}). The threshold may vary even up to $20$ DN for a given filtergram but is additionally modulated by a global trend.

By considering all these factors we can highlight the importance of individually extracting CHs without neglecting the local variations on CH size scales. Although manual input is needed, the extraction method implemented in CATCH aims to be as objective as possible without specifying any underlying extraction conditions except for the approach of the boundary gradient. 

\subsection{Distribution over CHs of Solar Cycle 24}
By analyzing the CHs of the SDO-era we not only gain a large sample of different CHs but also cover nearly one full Solar Cycle. As such, the sample includes CHs from the rising phase ($\approx 2010 / 2011$), the maximum phase ($\approx 2012 - 2014$), the decaying and minimum phase ($\approx 2015 - 2019$) of this cycle.

The CH parameters derived from the dataset are in good agreement with the study of \cite{2017hofmeister} who studied 288 low-latitude CHs near the maximum of Solar Cycle 24 and are as such a subset of this study. They found that the CH sizes are distributed around a median of $2.39  \times 10^{10}$ km$^{2}$ which is very close to the value derived in this study with a mean area of $(2.69 \pm 2.73)  \times 10^{10}$ km$^{2}$. Note here that the mean is strongly biased by the large amount of small CHs, of which a large portion is present in solar maximum. The spread in the CH sizes may also be influenced by the few large CHs ($5\%$ of CHs with an area exceeding $8 \times 10^{10}$ km$^{2}$). We excluded all polar and polar--connected CHs (as manually defined by the threshold tuning) from this statistical analysis which removes some of the largest CHs observed in this period from the study. This might be the reason why the extracted CH areas do not show the cycle dependence found by the Solar Cycle study by \cite{2017lowder}.

The mean signed magnetic field strength in our study shows a wider spread and higher average during the maximum phase than during the decaying and minimum phase. This was also previously stated by \cite{1982harvey}, who studied 33 CHs at 63 occasions and found that CHs near solar minima have  magnetic field strengths ranging from $1$ to $7$ G, while those detected near solar maxima, range from $3$ to $36$ G. In comparison, our values for the maximum ($3.4 \pm 2.1$ G) are significantly lower but for the minimum we are in good agreement. The difference may be due to the use of different instrumentation, as it has been shown that different instruments measure significantly different magnetic fields \citep[\textit{e.g.,}][]{2012liu}. Statistically, we find the mean absolute value of the mean magnetic field strength for all CHs under study to be $2.9 \pm 1.9$ G distributed from $0.4$ to $14.0$ G. Results from other studies are found inside this range ($\approx3$ G: \citealt{1978Bohlin}, $1-5$ G: \citealt{1989Obridko,2001Belenko}). Considering the property of the CHs open magnetic field configuration, we find that the flux balance, the ratio of the signed to the unsigned magnetic flux which is a measure of the percentage of open flux, is distributed from $10$ to $87\%$ which overlaps with the range found by \cite{2017hofmeister} of $6$ to $81\%$. A likely reason for the wide spread in the abundance of percentual open flux is that CHs of all evolutionary states are included in the dataset. The open magnetic field of the majority of CHs has been shown to possibly be due  to the mean magnetic field strength which varies with the evolution of a CH \citep[][]{2018heinemann_paperII}. This evolutionary process seems to be governed especially by interchange reconnection \citep{2004wang,2004Madjarska,2011krista,2014ma,2018kong} and flux emergence \citep{cranmer2009} and references therein). 

\cite{2017hofmeister}, \cite{2019hofmeister}, and \cite{2018heinemann_paperII} found that the abundance of the strong unipolar magnetic elements (flux tubes) is what defines the magnetic configuration of a CH. Notwithstanding that they cover only a small fraction of the CH area they contribute a major part of the total signed magnetic flux of the CH. \cite{2017hofmeister} found that strong FTs cover $1\%$ of the CHs area and contribute $38\%$ to the signed flux. These values are slightly lower than the ones we found in our study with $r_{A}=2.6\pm 1.8\%$ and $r_{\Phi}=58\pm 13\%$. This might be due to the differences in the extraction and definition of the strong FTs. Our results are in better agreement with the study of \cite{2018heinemann_paperII} who found values of $r_{A}~ \leqslant ~5\%$ and $r_{\Phi}=48$ to $71\%$. The recent study by \cite{2019hofmeister} found that these strong FTs have lifetimes larger that those of supergranular cells essentially making them the fundamental building blocks of CHs, and are not governed by the photospheric network motion.

\section{Summary}\label{s:sum}

In this comprehensive study we investigated the intensity gradient across the CH boundary to develop a new CH extraction method using an intensity-based threshold method as well as to estimate the uncertainties of the extracted CH boundaries. We successfully implemented the flexible and fast method into an easy-to-use GUI and applied it to the SDO-era to extract CHs. We created a CH catalogue of considerable size covering the time period from May 2010 to February 2019, which includes $707$ non-polar CHs that were closely analyzed. Our major findings can be summarized as follows:

\begin{enumerate}
    \item By incorporating the principle of the maximum gradient into the intensity-based threshold method we were able to:
    \begin{itemize}
        \item Create, for the first time, CH boundaries with reasonable estimates for the uncertainties
        \item Achieve a high consistency between boundaries extracted by different users
        \item Develop an objective as possible CH extraction method, without disregarding the advantages of manual user input 
    \end{itemize}
    \item Changes in the threshold due to small scale variations in the vicinity of CHs as well as global intensity variations as a consequence of the Solar Cycle show the importance of the individual extraction of a CH.
    \item By implementing the code into an SSW-IDL GUI we provide an user friendly environment for more objectively extracting CHs for scientific analysis, including reasonable uncertainties.
    \item Using CATCH we created an extensive catalogue for the CHs observed by SDO between its operational start in 2010 and February 2019. Over this era, we extracted and analyzed $707$ non-polar CHs and found them to exist in sizes ranging from $1.6 \times 10^{9}$ to $1.8 \times 10^{11}$ km$^{2}$. Small CHs ($<2 \times 10^{10}$ km$^{2}$) were found to be most abundant ($56\%$). The strength of the photospheric magnetic field underlying the CHs is distributed around $2.9\pm1.9$ G which is in agreement with most results found in literature and shows that CHs are mostly covered by low magnetic field.
    \item We confirm previous studies \citep[][]{2017hofmeister,2018heinemann_paperII, 2019hofmeister} that the magnetic configuration of CHs is highly dependent on the abundance and field strength of the small unipolar magnetic elements (flux tubes), that only cover a small fraction of the CH area.
\end{enumerate}

We plan to continue to develop CATCH. Planned major upgrades are the (partial--) implementation in Python and the option to use synoptic magnetograms. Also compatibility with Parker Solar Probe and Solar Orbiter are planned. New functionalities and upgrades will be published on GitHub and future user manual versions.


%

%

%
 \appendix   
 
\begin{figure} 
 \centerline{\includegraphics[width=1\textwidth,clip=]{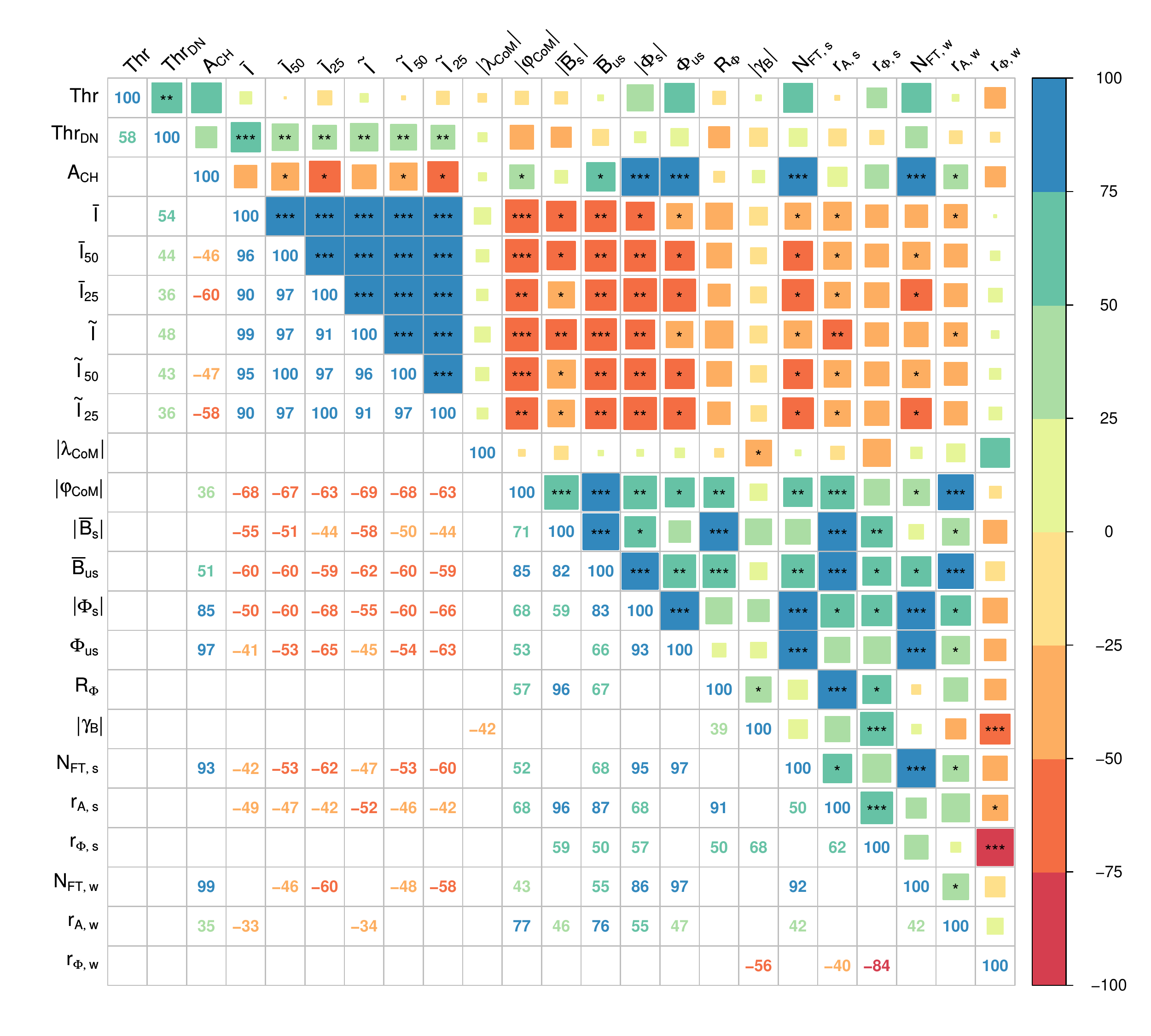}}
 \caption{Spearman correlation matrix of all the CH properties derived in the statistical analysis. The values, the square size as well as the color scheme represent the value of the spearman correlation coefficient which is given in percentual depiction. The black asterisks mark the significance level of the correlation: \**** indicates a  significance level of $p \le 0.001$, \*** indicates $p \le 0.01$ and \** indicates $p \le 0.05$. The values for all correlation coefficients with a significance level higher than $0.05$ were omitted. The parameters are listed in the Appendix.}\label{fig:corrmatrix}
 \end{figure}

Figure~\ref{fig:corrmatrix} shows the Spearman correlation coefficients of all the CH properties derived in the statistical analysis. The top right side shows representative squares for all correlation coefficients and additionally the significance level is marked with black asterisks (\**** indicates a  significance level of $p \le 0.001$, \*** indicates $p \le 0.01$ and \** indicates $p \le 0.05$). The left bottom side shows the values of the correlation coefficient with insignificant values (cc with $p > 0.05$) were omitted. The values have been converted to percent in order to improve visualization. Positive values correspond to a correlation and negative values to an anti--correlation. The parameters listed from left to right and top to bottom are the following: the optimal threshold as described in Section~\ref{subs:euv} in percent of the median solar disk intensity (Thr) and in DN (Thr$_{\mathrm{DN}}$); the CH area ($A_{\mathrm{CH}}$); the mean CH intensity ($\bar{I}$) and the mean intensities of the lowest $50\%$ and $25\%$ percentile of pixel intensities within the CH ($\bar{I}_{50}$, $\bar{I}_{25}$); the same for the median intensities ($\widetilde{I}$, $\widetilde{I}_{50}$, $\widetilde{I}_{25}$); the longitudinal ($|\lambda_{\mathrm{CoM}}|$) and latitudinal ($|\varphi_{\mathrm{CoM}}|$) position in absolute values; the absolute value of the signed mean magnetic field strength ($|\bar{B}_{\mathrm{s}}|$) and the unsigned mean magnetic field strength ($\bar{B}_{\mathrm{us}}$); the absolute value of the signed magnetic flux ($|\Phi_{\mathrm{s}}|$) and the unsigned magnetic flux ($\Phi_{\mathrm{us}}$); the flux balance ($R_{\Phi}$) and the absolute value of the skewness of the magnetic field ($|\gamma_{B}|$); the FT number, area ratio and flux ratio for both strong and weak FTs ($N_{\mathrm{FT},\mathrm{s}}$, $r_{A,\mathrm{s}}$, $r_{\Phi,\mathrm{s}}$, $N_{\mathrm{FT},\mathrm{w}}$, $r_{A,\mathrm{w}}$, $r_{\Phi,\mathrm{w}}$). 

%
 \begin{acks}

The SDO image data is available by courtesy of NASA and the respective science teams. This research has made use of the VizieR catalogue access tool, CDS, Strasbourg, France (DOI : 10.26093/cds/vizier). The original description of the VizieR service was published in A\&AS 143, 23. S.G.H., M.T., K.D., and A.M.V. acknowledge funding by the Austrian Space Applications Programme of the Austrian Research Promotion Agency FFG (859729, SWAMI, ASAP-11 4900217, CORDIM and ASAP-14 865972, SSCME). S.J.H. acknowledges support from the JungforscherInnenfonds der Steierm\"arkischen Sparkassen. S.G.H. thanks Evangelia Samara for providing the concept for the CATCH logo and Dr. Eleanna Asvestari for her incitement, which significantly enhanced the writing process. S.G.H. would also like to thank Aaron Hernandez-Perez for his input and support. A big thanks goes to all the testers of CATCH, who provided substantial feedback in improving it. We thank the anonymous referee for constructive comments, which helped to improve the manuscript and the tool. Disclosure of Potential Conflicts of Interest: The authors declare that they have no conflicts of interest.
 \end{acks}

%
%

\begin{thebibliography}{65}
\ifx\bisbn     \undefined \def\bisbn  #1{ISBN #1}\fi
\ifx\binits    \undefined \def\binits#1{#1}\fi
\ifx\bauthor   \undefined \def\bauthor#1{#1}\fi
\ifx\batitle   \undefined \def\batitle#1{#1}\fi
\ifx\bjtitle   \undefined \def\bjtitle#1{\textit{#1}}\fi
\ifx\bvolume   \undefined \def\bvolume#1{\textbf{#1}}\fi
\ifx\byear     \undefined \def\byear#1{#1}\fi
\ifx\bissue    \undefined \def\bissue#1{#1}\fi
\ifx\bfpage    \undefined \def\bfpage#1{#1}\fi
\ifx\blpage    \undefined \def\blpage #1{#1}\fi
\ifx\burl      \undefined \def\burl#1{\textsf{#1}}\fi
\ifx\href      \undefined \def\href#1#2{\textsf{#2}}\fi
\ifx\betal     \undefined \def\betal{\textit{et al.}}\fi
\ifx\bctitle   \undefined \def\bctitle#1{#1}\fi
\ifx\beditor   \undefined \def\beditor#1{#1}\fi
\ifx\bbtitle   \undefined \def\bbtitle#1{\textit{#1}}\fi
\ifx\bedition  \undefined \def\bedition#1{#1}\fi
\ifx\bseriesno \undefined \def\bseriesno#1{\textbf{#1}}\fi
\ifx\blocation \undefined \def\blocation#1{#1}\fi
\ifx\bsertitle \undefined \def\bsertitle#1{\textit{#1}}\fi
\ifx\bsnm      \undefined \def\bsnm#1{#1}\fi
\ifx\bsuffix   \undefined \def\bsuffix#1{#1}\fi
\ifx\bparticle \undefined \def\bparticle#1{#1}\fi
\ifx\barticle  \undefined \def\barticle#1{}\fi
\ifx\binstitute  \undefined \def\binstitute#1{#1}\fi
\ifx\bpublisher  \undefined \def\bpublisher#1{#1}\fi
\ifx\doiurl    \undefined \def\doiurl#1{\href{#1}{\textsf{DOI}}}\fi
\makeatletter
\def\safeHref#1#2#3{\in@{http}{#2}\ifin@\href{#2}{#3}\else\href{#1#2}{#3}\fi}
\makeatother
\ifx\adsurl    \undefined
  \def\adsurl#1{\safeHref{http://adsabs.harvard.edu/abs/}{#1}{\textsf{ADS}}}\fi
\ifx\arxivurl  \undefined
  \def\arxivurl#1{\safeHref{http://arxiv.org/abs/}{#1}{\textsf{arXiv}}}\fi
\ifx\botherref \undefined \def\botherref#1{}\fi
\ifx\url       \undefined \def\url#1{\textsf{#1}}\fi
\ifx\bchapter  \undefined \def\bchapter#1{}\fi
\ifx\bbook     \undefined \def\bbook#1{}\fi
\ifx\bcomment  \undefined \def\bcomment#1{#1}\fi
\ifx\oauthor   \undefined \def\oauthor#1{#1}\fi
\ifx\citeauthoryear \undefined\def \citeauthoryear#1{#1}\fi
\def\endbibitem {}
\ifx\bconflocation  \undefined \def\bconflocation#1{#1} \fi

\bibitem[\protect\citeauthoryear{{Altschuler} and
  {Newkirk}}{1969}]{1969altschuler_PFSS}
\begin{barticle}
\bauthor{\bsnm{{Altschuler}}, \binits{M.D.}},
\bauthor{\bsnm{{Newkirk}}, \binits{G.}}:
\byear{1969},
\batitle{{Magnetic Fields and the Structure of the Solar Corona. I: Methods of
  Calculating Coronal Fields}}.
\bjtitle{\solphys}
\bvolume{9}(\bissue{1}),
\bfpage{131}.
\doiurl{https://doi.org/10.1007/BF00145734}.
\adsurl{https://ui.adsabs.harvard.edu/abs/1969SoPh....9..131A}.
\end{barticle}
\endbibitem

\bibitem[\protect\citeauthoryear{Arge and Pizzo}{2000}]{2000arge}
\begin{barticle}
\bauthor{\bsnm{Arge}, \binits{C.N.}},
\bauthor{\bsnm{Pizzo}, \binits{V.J.}}:
\byear{2000},
\batitle{Improvement in the prediction of solar wind conditions using near-real
  time solar magnetic field updates}.
\bjtitle{Journal of Geophysical Research}
\bvolume{105},
\bfpage{10465}.
\doiurl{https://doi.org/10.1029/1999JA000262}.
\burl{http://adsabs.harvard.edu/abs/2000JGR...10510465A}.
\end{barticle}
\endbibitem

\bibitem[\protect\citeauthoryear{{Asvestari}
  \textit{et~al.}}{2019}]{2019asvestari}
\begin{botherref}
\oauthor{\bsnm{{Asvestari}}, \binits{E.}},
\oauthor{\bsnm{G.}, \binits{H.S.}},
\oauthor{\bsnm{{Temmer}}, \binits{M.}},
\oauthor{\bsnm{{Pomoell}}, \binits{J.}},
\oauthor{\bsnm{{Kilpua}}, \binits{E.}},
\oauthor{\bsnm{{Magdalenic}}, \binits{J.}},
\oauthor{\bsnm{{Poedts}}, \binits{S.}}:
2019,
{}.
\textit{\jgr}.
\doiurl{submitted}.
\end{botherref}
\endbibitem

\bibitem[\protect\citeauthoryear{{Belenko}}{2001}]{2001Belenko}
\begin{barticle}
\bauthor{\bsnm{{Belenko}}, \binits{I.A.}}:
\byear{2001},
\batitle{{Coronal Hole Evolution During 1996-1999}}.
\bjtitle{\solphys}
\bvolume{199}(\bissue{1}),
\bfpage{23}.
\doiurl{https://doi.org/10.1023/A:1010372926629}.
\adsurl{https://ui.adsabs.harvard.edu/abs/2001SoPh..199...23B}.
\end{barticle}
\endbibitem

\bibitem[\protect\citeauthoryear{{Bohlin} and {Sheeley}}{1978}]{1978Bohlin}
\begin{barticle}
\bauthor{\bsnm{{Bohlin}}, \binits{J.D.}},
\bauthor{\bsnm{{Sheeley}}, \binits{N.R.} \bsuffix{Jr.}}:
\byear{1978},
\batitle{{Extreme ultraviolet observations of coronal holes. II - Association
  of holes with solar magnetic fields and a model for their formation during
  the solar cycle}}.
\bjtitle{\solphys}
\bvolume{56},
\bfpage{125}.
\doiurl{https://doi.org/10.1007/BF00152639}.
\adsurl{https://ui.adsabs.harvard.edu/abs/1978SoPh...56..125B}.
\end{barticle}
\endbibitem

\bibitem[\protect\citeauthoryear{Boucheron, Valluri, and
  McAteer}{2016}]{Boucheron2016}
\begin{barticle}
\bauthor{\bsnm{Boucheron}, \binits{L.E.}},
\bauthor{\bsnm{Valluri}, \binits{M.}},
\bauthor{\bsnm{McAteer}, \binits{R.T.J.}}:
\byear{2016},
\batitle{Segmentation of coronal holes using active contours without edges}.
\bjtitle{Solar Physics}
\bvolume{291}(\bissue{8}),
\bfpage{2353}.
\doiurl{https://doi.org/10.1007/s11207-016-0985-z}.
\burl{https://doi.org/10.1007/s11207-016-0985-z}.
\end{barticle}
\endbibitem

\bibitem[\protect\citeauthoryear{{Caplan}, {Downs}, and
  {Linker}}{2016}]{2016caplan_LBC_IIT}
\begin{barticle}
\bauthor{\bsnm{{Caplan}}, \binits{R.M.}},
\bauthor{\bsnm{{Downs}}, \binits{C.}},
\bauthor{\bsnm{{Linker}}, \binits{J.A.}}:
\byear{2016},
\batitle{{Synchronic Coronal Hole Mapping Using Multi-instrument EUV Images:
  Data Preparation and Detection Method}}.
\bjtitle{\apj}
\bvolume{823},
\bfpage{53}.
\doiurl{https://doi.org/10.3847/0004-637X/823/1/53}.
\adsurl{2016ApJ...823...53C}.
\end{barticle}
\endbibitem

\bibitem[\protect\citeauthoryear{{Couvidat}
  \textit{et~al.}}{2016}]{2016couvidat_HMI}
\begin{barticle}
\bauthor{\bsnm{{Couvidat}}, \binits{S.}},
\bauthor{\bsnm{{Schou}}, \binits{J.}},
\bauthor{\bsnm{{Hoeksema}}, \binits{J.T.}},
\bauthor{\bsnm{{Bogart}}, \binits{R.S.}},
\bauthor{\bsnm{{Bush}}, \binits{R.I.}},
\bauthor{\bsnm{{Duvall}}, \binits{T.L.}},
\bauthor{\bsnm{{Liu}}, \binits{Y.}},
\bauthor{\bsnm{{Norton}}, \binits{A.A.}},
\bauthor{\bsnm{{Scherrer}}, \binits{P.H.}}:
\byear{2016},
\batitle{{Observables Processing for the Helioseismic and Magnetic Imager
  Instrument on the Solar Dynamics Observatory}}.
\bjtitle{\solphys}
\bvolume{291},
\bfpage{1887}.
\doiurl{https://doi.org/10.1007/s11207-016-0957-3}.
\adsurl{2016SoPh..291.1887C}.
\end{barticle}
\endbibitem

\bibitem[\protect\citeauthoryear{{Cranmer}}{2002}]{cranmer2002}
\begin{barticle}
\bauthor{\bsnm{{Cranmer}}, \binits{S.R.}}:
\byear{2002},
\batitle{{Coronal Holes and the High-Speed Solar Wind}}.
\bjtitle{\ssr}
\bvolume{101},
\bfpage{229}.
\adsurl{2002SSRv..101..229C}.
\end{barticle}
\endbibitem

\bibitem[\protect\citeauthoryear{{Cranmer}}{2009}]{cranmer2009}
\begin{barticle}
\bauthor{\bsnm{{Cranmer}}, \binits{S.R.}}:
\byear{2009},
\batitle{{Coronal Holes}}.
\bjtitle{Living Reviews in Solar Physics}
\bvolume{6},
\bfpage{3}.
\doiurl{https://doi.org/10.12942/lrsp-2009-3}.
\adsurl{2009LRSP....6....3C}.
\end{barticle}
\endbibitem

\bibitem[\protect\citeauthoryear{{Delaboudini{\`e}re}
  \textit{et~al.}}{1995}]{1995delaboudiniere-EIT}
\begin{barticle}
\bauthor{\bsnm{{Delaboudini{\`e}re}}, \binits{J.-P.}},
\bauthor{\bsnm{{Artzner}}, \binits{G.E.}},
\bauthor{\bsnm{{Brunaud}}, \binits{J.}},
\bauthor{\bsnm{{Gabriel}}, \binits{A.H.}},
\bauthor{\bsnm{{Hochedez}}, \binits{J.F.}},
\bauthor{\bsnm{{Millier}}, \binits{F.}},
\bauthor{\bsnm{{Song}}, \binits{X.Y.}},
\bauthor{\bsnm{{Au}}, \binits{B.}},
\bauthor{\bsnm{{Dere}}, \binits{K.P.}},
\bauthor{\bsnm{{Howard}}, \binits{R.A.}},
\bauthor{\bsnm{{Kreplin}}, \binits{R.}},
\bauthor{\bsnm{{Michels}}, \binits{D.J.}},
\bauthor{\bsnm{{Moses}}, \binits{J.D.}},
\bauthor{\bsnm{{Defise}}, \binits{J.M.}},
\bauthor{\bsnm{{Jamar}}, \binits{C.}},
\bauthor{\bsnm{{Rochus}}, \binits{P.}},
\bauthor{\bsnm{{Chauvineau}}, \binits{J.P.}},
\bauthor{\bsnm{{Marioge}}, \binits{J.P.}},
\bauthor{\bsnm{{Catura}}, \binits{R.C.}},
\bauthor{\bsnm{{Lemen}}, \binits{J.R.}},
\bauthor{\bsnm{{Shing}}, \binits{L.}},
\bauthor{\bsnm{{Stern}}, \binits{R.A.}},
\bauthor{\bsnm{{Gurman}}, \binits{J.B.}},
\bauthor{\bsnm{{Neupert}}, \binits{W.M.}},
\bauthor{\bsnm{{Maucherat}}, \binits{A.}},
\bauthor{\bsnm{{Clette}}, \binits{F.}},
\bauthor{\bsnm{{Cugnon}}, \binits{P.}},
\bauthor{\bsnm{{van Dessel}}, \binits{E.L.}}:
\byear{1995},
\batitle{{EIT: Extreme-Ultraviolet Imaging Telescope for the SOHO Mission}}.
\bjtitle{\solphys}
\bvolume{162},
\bfpage{291}.
\doiurl{https://doi.org/10.1007/BF00733432}.
\adsurl{https://ui.adsabs.harvard.edu/abs/1995SoPh..162..291D}.
\end{barticle}
\endbibitem

\bibitem[\protect\citeauthoryear{Delouille
  \textit{et~al.}}{2018}]{2018DELOUILLE}
\begin{bchapter}
\bauthor{\bsnm{Delouille}, \binits{V.}},
\bauthor{\bsnm{Hofmeister}, \binits{S.J.}},
\bauthor{\bsnm{Reiss}, \binits{M.A.}},
\bauthor{\bsnm{Mampaey}, \binits{B.}},
\bauthor{\bsnm{Temmer}, \binits{M.}},
\bauthor{\bsnm{Veronig}, \binits{A.}}:
\byear{2018},
\bctitle{Chapter 15 - coronal holes detection using supervised classification}.
In: \beditor{\bsnm{Camporeale}, \binits{E.}},
\beditor{\bsnm{Wing}, \binits{S.}},
\beditor{\bsnm{Johnson}, \binits{J.R.}} (eds.)
\bbtitle{Machine Learning Techniques for Space Weather},
\bpublisher{Elsevier}, \blocation{???},
\bfpage{365 }.
\bisbn{978-0-12-811788-0}.
\doiurl{https://doi.org/10.1016/B978-0-12-811788-0.00015-9}.
\burl{http://www.sciencedirect.com/science/article/pii/B9780128117880000159}.
\end{bchapter}
\endbibitem

\bibitem[\protect\citeauthoryear{{Domingo}, {Fleck}, and
  {Poland}}{1995}]{1995soho}
\begin{barticle}
\bauthor{\bsnm{{Domingo}}, \binits{V.}},
\bauthor{\bsnm{{Fleck}}, \binits{B.}},
\bauthor{\bsnm{{Poland}}, \binits{A.I.}}:
\byear{1995},
\batitle{{The SOHO Mission: an Overview}}.
\bjtitle{\solphys}
\bvolume{162},
\bfpage{1}.
\doiurl{https://doi.org/10.1007/BF00733425}.
\adsurl{http://cdsads.u-strasbg.fr/abs/1995SoPh..162....1D}.
\end{barticle}
\endbibitem

\bibitem[\protect\citeauthoryear{{Edmondson}
  \textit{et~al.}}{2010}]{2010edmondson}
\begin{barticle}
\bauthor{\bsnm{{Edmondson}}, \binits{J.K.}},
\bauthor{\bsnm{{Antiochos}}, \binits{S.K.}},
\bauthor{\bsnm{{DeVore}}, \binits{C.R.}},
\bauthor{\bsnm{{Lynch}}, \binits{B.J.}},
\bauthor{\bsnm{{Zurbuchen}}, \binits{T.H.}}:
\byear{2010},
\batitle{{Interchange Reconnection and Coronal Hole Dynamics}}.
\bjtitle{\apj}
\bvolume{714},
\bfpage{517}.
\doiurl{https://doi.org/10.1088/0004-637X/714/1/517}.
\adsurl{https://ui.adsabs.harvard.edu/abs/2010ApJ...714..517E}.
\end{barticle}
\endbibitem

\bibitem[\protect\citeauthoryear{{Garton}, {Gallagher}, and
  {Murray}}{2018}]{2018garton}
\begin{barticle}
\bauthor{\bsnm{{Garton}}, \binits{T.M.}},
\bauthor{\bsnm{{Gallagher}}, \binits{P.T.}},
\bauthor{\bsnm{{Murray}}, \binits{S.A.}}:
\byear{2018},
\batitle{{Automated coronal hole identification via multi-thermal intensity
  segmentation}}.
\bjtitle{Journal of Space Weather and Space Climate}
\bvolume{8}(\bissue{27}),
\bfpage{A02}.
\doiurl{https://doi.org/10.1051/swsc/2017039}.
\adsurl{2018JSWSC...8A...2G}.
\end{barticle}
\endbibitem

\bibitem[\protect\citeauthoryear{{Hahn}, {Landi}, and {Savin}}{2011}]{2011hahn}
\begin{barticle}
\bauthor{\bsnm{{Hahn}}, \binits{M.}},
\bauthor{\bsnm{{Landi}}, \binits{E.}},
\bauthor{\bsnm{{Savin}}, \binits{D.W.}}:
\byear{2011},
\batitle{{Differential Emission Measure Analysis of a Polar Coronal Hole during
  the Solar Minimum in 2007}}.
\bjtitle{\apj}
\bvolume{736}(\bissue{2}),
\bfpage{101}.
\doiurl{https://doi.org/10.1088/0004-637X/736/2/101}.
\adsurl{https://ui.adsabs.harvard.edu/abs/2011ApJ...736..101H}.
\end{barticle}
\endbibitem

\bibitem[\protect\citeauthoryear{{Hamada} \textit{et~al.}}{2018}]{2018hamada}
\begin{barticle}
\bauthor{\bsnm{{Hamada}}, \binits{A.}},
\bauthor{\bsnm{{Asikainen}}, \binits{T.}},
\bauthor{\bsnm{{Virtanen}}, \binits{I.}},
\bauthor{\bsnm{{Mursula}}, \binits{K.}}:
\byear{2018},
\batitle{{Automated Identification of Coronal Holes from Synoptic EUV Maps}}.
\bjtitle{\solphys}
\bvolume{293},
\bfpage{71}.
\doiurl{https://doi.org/10.1007/s11207-018-1289-2}.
\adsurl{https://ui.adsabs.harvard.edu/abs/2018SoPh..293...71H}.
\end{barticle}
\endbibitem

\bibitem[\protect\citeauthoryear{{Harvey}, {Sheeley}, and
  {Harvey}}{1982}]{1982harvey}
\begin{barticle}
\bauthor{\bsnm{{Harvey}}, \binits{K.L.}},
\bauthor{\bsnm{{Sheeley}}, \binits{N.R.} \bsuffix{Jr.}},
\bauthor{\bsnm{{Harvey}}, \binits{J.W.}}:
\byear{1982},
\batitle{{Magnetic measurements of coronal holes during 1975-1980}}.
\bjtitle{\solphys}
\bvolume{79},
\bfpage{149}.
\doiurl{https://doi.org/10.1007/BF00146979}.
\adsurl{https://ui.adsabs.harvard.edu/abs/1982SoPh...79..149H}.
\end{barticle}
\endbibitem

\bibitem[\protect\citeauthoryear{{Heinemann}
  \textit{et~al.}}{2018a}]{2018heinemann_paperI}
\begin{barticle}
\bauthor{\bsnm{{Heinemann}}, \binits{S.G.}},
\bauthor{\bsnm{{Temmer}}, \binits{M.}},
\bauthor{\bsnm{{Hofmeister}}, \binits{S.J.}},
\bauthor{\bsnm{{Veronig}}, \binits{A.M.}},
\bauthor{\bsnm{{Vennerstr{\o}m}}, \binits{S.}}:
\byear{2018}a,
\batitle{{Three-phase Evolution of a Coronal Hole. I. 360$^{\circ}$ Remote
  Sensing and In Situ Observations}}.
\bjtitle{\apj}
\bvolume{861},
\bfpage{151}.
\doiurl{https://doi.org/10.3847/1538-4357/aac897}.
\adsurl{2018ApJ...861..151H}.
\end{barticle}
\endbibitem

\bibitem[\protect\citeauthoryear{{Heinemann}
  \textit{et~al.}}{2018b}]{2018heinemann_paperII}
\begin{barticle}
\bauthor{\bsnm{{Heinemann}}, \binits{S.G.}},
\bauthor{\bsnm{{Hofmeister}}, \binits{S.J.}},
\bauthor{\bsnm{{Veronig}}, \binits{A.M.}},
\bauthor{\bsnm{{Temmer}}, \binits{M.}}:
\byear{2018}b,
\batitle{{Three-phase Evolution of a Coronal Hole. II. The Magnetic Field}}.
\bjtitle{\apj}
\bvolume{863},
\bfpage{29}.
\doiurl{https://doi.org/10.3847/1538-4357/aad095}.
\adsurl{2018ApJ...863...29H}.
\end{barticle}
\endbibitem

\bibitem[\protect\citeauthoryear{{Hofmeister}
  \textit{et~al.}}{2017}]{2017hofmeister}
\begin{barticle}
\bauthor{\bsnm{{Hofmeister}}, \binits{S.J.}},
\bauthor{\bsnm{{Veronig}}, \binits{A.}},
\bauthor{\bsnm{{Reiss}}, \binits{M.A.}},
\bauthor{\bsnm{{Temmer}}, \binits{M.}},
\bauthor{\bsnm{{Vennerstrom}}, \binits{S.}},
\bauthor{\bsnm{{Vr{\v s}nak}}, \binits{B.}},
\bauthor{\bsnm{{Heber}}, \binits{B.}}:
\byear{2017},
\batitle{{Characteristics of Low-latitude Coronal Holes near the Maximum of
  Solar Cycle 24}}.
\bjtitle{\apj}
\bvolume{835},
\bfpage{268}.
\doiurl{https://doi.org/10.3847/1538-4357/835/2/268}.
\adsurl{2017ApJ...835..268H}.
\end{barticle}
\endbibitem

\bibitem[\protect\citeauthoryear{{Hofmeister}
  \textit{et~al.}}{2018}]{2018hofmeister}
\begin{barticle}
\bauthor{\bsnm{{Hofmeister}}, \binits{S.J.}},
\bauthor{\bsnm{{Veronig}}, \binits{A.}},
\bauthor{\bsnm{{Temmer}}, \binits{M.}},
\bauthor{\bsnm{{Vennerstrom}}, \binits{S.}},
\bauthor{\bsnm{{Heber}}, \binits{B.}},
\bauthor{\bsnm{{Vr{\v s}nak}}, \binits{B.}}:
\byear{2018},
\batitle{{The Dependence of the Peak Velocity of High-Speed Solar Wind Streams
  as Measured in the Ecliptic by ACE and the STEREO satellites on the Area and
  Co-latitude of Their Solar Source Coronal Holes}}.
\bjtitle{Journal of Geophysical Research (Space Physics)}
\bvolume{123},
\bfpage{1738}.
\doiurl{https://doi.org/10.1002/2017JA024586}.
\adsurl{https://ui.adsabs.harvard.edu/abs/2018JGRA..123.1738H}.
\end{barticle}
\endbibitem

\bibitem[\protect\citeauthoryear{{Hofmeister}
  \textit{et~al.}}{2019}]{2019hofmeister}
\begin{botherref}
\oauthor{\bsnm{{Hofmeister}}, \binits{S.J.}},
\oauthor{\bsnm{{Utz}}, \binits{D.}},
\oauthor{\bsnm{{Heinemann}}, \binits{S.G.}},
\oauthor{\bsnm{{Veronig}}, \binits{A.M.}},
\oauthor{\bsnm{{Temmer}}, \binits{M.}}:
2019,
{The Magnetic Structure of Coronal Holes}.
\textit{\aap}.
\doiurl{https://doi.org/under review, minor comments}.
\end{botherref}
\endbibitem

\bibitem[\protect\citeauthoryear{{Howard}
  \textit{et~al.}}{2008}]{2008howard_SECCHI}
\begin{barticle}
\bauthor{\bsnm{{Howard}}, \binits{R.A.}},
\bauthor{\bsnm{{Moses}}, \binits{J.D.}},
\bauthor{\bsnm{{Vourlidas}}, \binits{A.}},
\bauthor{\bsnm{{Newmark}}, \binits{J.S.}},
\bauthor{\bsnm{{Socker}}, \binits{D.G.}},
\bauthor{\bsnm{{Plunkett}}, \binits{S.P.}},
\bauthor{\bsnm{{Korendyke}}, \binits{C.M.}},
\bauthor{\bsnm{{Cook}}, \binits{J.W.}},
\bauthor{\bsnm{{Hurley}}, \binits{A.}},
\bauthor{\bsnm{{Davila}}, \binits{J.M.}},
\bauthor{\bsnm{{Thompson}}, \binits{W.T.}},
\bauthor{\bsnm{{St Cyr}}, \binits{O.C.}},
\bauthor{\bsnm{{Mentzell}}, \binits{E.}},
\bauthor{\bsnm{{Mehalick}}, \binits{K.}},
\bauthor{\bsnm{{Lemen}}, \binits{J.R.}},
\bauthor{\bsnm{{Wuelser}}, \binits{J.P.}},
\bauthor{\bsnm{{Duncan}}, \binits{D.W.}},
\bauthor{\bsnm{{Tarbell}}, \binits{T.D.}},
\bauthor{\bsnm{{Wolfson}}, \binits{C.J.}},
\bauthor{\bsnm{{Moore}}, \binits{A.}},
\bauthor{\bsnm{{Harrison}}, \binits{R.A.}},
\bauthor{\bsnm{{Waltham}}, \binits{N.R.}},
\bauthor{\bsnm{{Lang}}, \binits{J.}},
\bauthor{\bsnm{{Davis}}, \binits{C.J.}},
\bauthor{\bsnm{{Eyles}}, \binits{C.J.}},
\bauthor{\bsnm{{Mapson-Menard}}, \binits{H.}},
\bauthor{\bsnm{{Simnett}}, \binits{G.M.}},
\bauthor{\bsnm{{Halain}}, \binits{J.P.}},
\bauthor{\bsnm{{Defise}}, \binits{J.M.}},
\bauthor{\bsnm{{Mazy}}, \binits{E.}},
\bauthor{\bsnm{{Rochus}}, \binits{P.}},
\bauthor{\bsnm{{Mercier}}, \binits{R.}},
\bauthor{\bsnm{{Ravet}}, \binits{M.F.}},
\bauthor{\bsnm{{Delmotte}}, \binits{F.}},
\bauthor{\bsnm{{Auchere}}, \binits{F.}},
\bauthor{\bsnm{{Delaboudiniere}}, \binits{J.P.}},
\bauthor{\bsnm{{Bothmer}}, \binits{V.}},
\bauthor{\bsnm{{Deutsch}}, \binits{W.}},
\bauthor{\bsnm{{Wang}}, \binits{D.}},
\bauthor{\bsnm{{Rich}}, \binits{N.}},
\bauthor{\bsnm{{Cooper}}, \binits{S.}},
\bauthor{\bsnm{{Stephens}}, \binits{V.}},
\bauthor{\bsnm{{Maahs}}, \binits{G.}},
\bauthor{\bsnm{{Baugh}}, \binits{R.}},
\bauthor{\bsnm{{McMullin}}, \binits{D.}},
\bauthor{\bsnm{{Carter}}, \binits{T.}}:
\byear{2008},
\batitle{{Sun Earth Connection Coronal and Heliospheric Investigation
  (SECCHI)}}.
\bjtitle{\ssr}
\bvolume{136},
\bfpage{67}.
\doiurl{https://doi.org/10.1007/s11214-008-9341-4}.
\adsurl{2008SSRv..136...67H}.
\end{barticle}
\endbibitem

\bibitem[\protect\citeauthoryear{{Huang}, {Lin}, and {Lee}}{2019}]{2019huang}
\begin{barticle}
\bauthor{\bsnm{{Huang}}, \binits{G.-H.}},
\bauthor{\bsnm{{Lin}}, \binits{C.-H.}},
\bauthor{\bsnm{{Lee}}, \binits{L.-C.}}:
\byear{2019},
\batitle{{Examination of the EUV Intensity in the Open Magnetic Field Regions
  Associated with Coronal Holes}}.
\bjtitle{\apj}
\bvolume{874},
\bfpage{45}.
\doiurl{https://doi.org/10.3847/1538-4357/ab06f0}.
\adsurl{2019ApJ...874...45H}.
\end{barticle}
\endbibitem

\bibitem[\protect\citeauthoryear{{Illarionov} and
  {Tlatov}}{2018}]{2018illarionov}
\begin{barticle}
\bauthor{\bsnm{{Illarionov}}, \binits{E.A.}},
\bauthor{\bsnm{{Tlatov}}, \binits{A.G.}}:
\byear{2018},
\batitle{{Segmentation of coronal holes in solar disc images with a
  convolutional neural network}}.
\bjtitle{\mnras}
\bvolume{481},
\bfpage{5014}.
\doiurl{https://doi.org/10.1093/mnras/sty2628}.
\adsurl{https://ui.adsabs.harvard.edu/abs/2018MNRAS.481.5014I}.
\end{barticle}
\endbibitem

\bibitem[\protect\citeauthoryear{{Kaiser}
  \textit{et~al.}}{2008}]{2008kaiser_STEREO}
\begin{barticle}
\bauthor{\bsnm{{Kaiser}}, \binits{M.L.}},
\bauthor{\bsnm{{Kucera}}, \binits{T.A.}},
\bauthor{\bsnm{{Davila}}, \binits{J.M.}},
\bauthor{\bsnm{{St.~Cyr}}, \binits{O.C.}},
\bauthor{\bsnm{{Guhathakurta}}, \binits{M.}},
\bauthor{\bsnm{{Christian}}, \binits{E.}}:
\byear{2008},
\batitle{{The STEREO Mission: An Introduction}}.
\bjtitle{\ssr}
\bvolume{136},
\bfpage{5}.
\doiurl{https://doi.org/10.1007/s11214-007-9277-0}.
\adsurl{2008SSRv..136....5K}.
\end{barticle}
\endbibitem

\bibitem[\protect\citeauthoryear{{Karachik}, {Pevtsov}, and
  {Abramenko}}{2010}]{2010Karachik}
\begin{barticle}
\bauthor{\bsnm{{Karachik}}, \binits{N.V.}},
\bauthor{\bsnm{{Pevtsov}}, \binits{A.A.}},
\bauthor{\bsnm{{Abramenko}}, \binits{V.I.}}:
\byear{2010},
\batitle{{Formation of Coronal Holes on the Ashes of Active Regions}}.
\bjtitle{\apj}
\bvolume{714}(\bissue{2}),
\bfpage{1672}.
\doiurl{https://doi.org/10.1088/0004-637X/714/2/1672}.
\adsurl{https://ui.adsabs.harvard.edu/abs/2010ApJ...714.1672K}.
\end{barticle}
\endbibitem

\bibitem[\protect\citeauthoryear{{Kong} \textit{et~al.}}{2018}]{2018kong}
\begin{barticle}
\bauthor{\bsnm{{Kong}}, \binits{D.F.}},
\bauthor{\bsnm{{Pan}}, \binits{G.M.}},
\bauthor{\bsnm{{Yan}}, \binits{X.L.}},
\bauthor{\bsnm{{Wang}}, \binits{J.C.}},
\bauthor{\bsnm{{Li}}, \binits{Q.L.}}:
\byear{2018},
\batitle{{Observational Evidence of Interchange Reconnection between a Solar
  Coronal Hole and a Small Emerging Active Region}}.
\bjtitle{\apj}
\bvolume{863}(\bissue{2}),
\bfpage{L22}.
\doiurl{https://doi.org/10.3847/2041-8213/aad777}.
\adsurl{https://ui.adsabs.harvard.edu/abs/2018ApJ...863L..22K}.
\end{barticle}
\endbibitem

\bibitem[\protect\citeauthoryear{{Krista} and {Gallagher}}{2009}]{2009krista}
\begin{barticle}
\bauthor{\bsnm{{Krista}}, \binits{L.D.}},
\bauthor{\bsnm{{Gallagher}}, \binits{P.T.}}:
\byear{2009},
\batitle{{Automated Coronal Hole Detection Using Local Intensity Thresholding
  Techniques}}.
\bjtitle{\solphys}
\bvolume{256},
\bfpage{87}.
\doiurl{https://doi.org/10.1007/s11207-009-9357-2}.
\adsurl{2009SoPh..256...87K}.
\end{barticle}
\endbibitem

\bibitem[\protect\citeauthoryear{{Krista}, {Gallagher}, and
  {Bloomfield}}{2011}]{2011krista}
\begin{barticle}
\bauthor{\bsnm{{Krista}}, \binits{L.D.}},
\bauthor{\bsnm{{Gallagher}}, \binits{P.T.}},
\bauthor{\bsnm{{Bloomfield}}, \binits{D.S.}}:
\byear{2011},
\batitle{{Short-term Evolution of Coronal Hole Boundaries}}.
\bjtitle{\apj}
\bvolume{731}(\bissue{2}),
\bfpage{L26}.
\doiurl{https://doi.org/10.1088/2041-8205/731/2/L26}.
\adsurl{https://ui.adsabs.harvard.edu/abs/2011ApJ...731L..26K}.
\end{barticle}
\endbibitem

\bibitem[\protect\citeauthoryear{{Lemen} \textit{et~al.}}{2012}]{2012lemen_AIA}
\begin{barticle}
\bauthor{\bsnm{{Lemen}}, \binits{J.R.}},
\bauthor{\bsnm{{Title}}, \binits{A.M.}},
\bauthor{\bsnm{{Akin}}, \binits{D.J.}},
\bauthor{\bsnm{{Boerner}}, \binits{P.F.}},
\bauthor{\bsnm{{Chou}}, \binits{C.}},
\bauthor{\bsnm{{Drake}}, \binits{J.F.}},
\bauthor{\bsnm{{Duncan}}, \binits{D.W.}},
\bauthor{\bsnm{{Edwards}}, \binits{C.G.}},
\bauthor{\bsnm{{Friedlaender}}, \binits{F.M.}},
\bauthor{\bsnm{{Heyman}}, \binits{G.F.}},
\bauthor{\bsnm{{Hurlburt}}, \binits{N.E.}},
\bauthor{\bsnm{{Katz}}, \binits{N.L.}},
\bauthor{\bsnm{{Kushner}}, \binits{G.D.}},
\bauthor{\bsnm{{Levay}}, \binits{M.}},
\bauthor{\bsnm{{Lindgren}}, \binits{R.W.}},
\bauthor{\bsnm{{Mathur}}, \binits{D.P.}},
\bauthor{\bsnm{{McFeaters}}, \binits{E.L.}},
\bauthor{\bsnm{{Mitchell}}, \binits{S.}},
\bauthor{\bsnm{{Rehse}}, \binits{R.A.}},
\bauthor{\bsnm{{Schrijver}}, \binits{C.J.}},
\bauthor{\bsnm{{Springer}}, \binits{L.A.}},
\bauthor{\bsnm{{Stern}}, \binits{R.A.}},
\bauthor{\bsnm{{Tarbell}}, \binits{T.D.}},
\bauthor{\bsnm{{Wuelser}}, \binits{J.-P.}},
\bauthor{\bsnm{{Wolfson}}, \binits{C.J.}},
\bauthor{\bsnm{{Yanari}}, \binits{C.}},
\bauthor{\bsnm{{Bookbinder}}, \binits{J.A.}},
\bauthor{\bsnm{{Cheimets}}, \binits{P.N.}},
\bauthor{\bsnm{{Caldwell}}, \binits{D.}},
\bauthor{\bsnm{{Deluca}}, \binits{E.E.}},
\bauthor{\bsnm{{Gates}}, \binits{R.}},
\bauthor{\bsnm{{Golub}}, \binits{L.}},
\bauthor{\bsnm{{Park}}, \binits{S.}},
\bauthor{\bsnm{{Podgorski}}, \binits{W.A.}},
\bauthor{\bsnm{{Bush}}, \binits{R.I.}},
\bauthor{\bsnm{{Scherrer}}, \binits{P.H.}},
\bauthor{\bsnm{{Gummin}}, \binits{M.A.}},
\bauthor{\bsnm{{Smith}}, \binits{P.}},
\bauthor{\bsnm{{Auker}}, \binits{G.}},
\bauthor{\bsnm{{Jerram}}, \binits{P.}},
\bauthor{\bsnm{{Pool}}, \binits{P.}},
\bauthor{\bsnm{{Soufli}}, \binits{R.}},
\bauthor{\bsnm{{Windt}}, \binits{D.L.}},
\bauthor{\bsnm{{Beardsley}}, \binits{S.}},
\bauthor{\bsnm{{Clapp}}, \binits{M.}},
\bauthor{\bsnm{{Lang}}, \binits{J.}},
\bauthor{\bsnm{{Waltham}}, \binits{N.}}:
\byear{2012},
\batitle{{The Atmospheric Imaging Assembly (AIA) on the Solar Dynamics
  Observatory (SDO)}}.
\bjtitle{\solphys}
\bvolume{275},
\bfpage{17}.
\doiurl{https://doi.org/10.1007/s11207-011-9776-8}.
\adsurl{2012SoPh..275...17L}.
\end{barticle}
\endbibitem

\bibitem[\protect\citeauthoryear{Linker \textit{et~al.}}{2017}]{Linker_2017}
\begin{barticle}
\bauthor{\bsnm{Linker}, \binits{J.A.}},
\bauthor{\bsnm{Caplan}, \binits{R.M.}},
\bauthor{\bsnm{Downs}, \binits{C.}},
\bauthor{\bsnm{Riley}, \binits{P.}},
\bauthor{\bsnm{Mikic}, \binits{Z.}},
\bauthor{\bsnm{Lionello}, \binits{R.}},
\bauthor{\bsnm{Henney}, \binits{C.J.}},
\bauthor{\bsnm{Arge}, \binits{C.N.}},
\bauthor{\bsnm{Liu}, \binits{Y.}},
\bauthor{\bsnm{Derosa}, \binits{M.L.}},
\bauthor{\bsnm{Yeates}, \binits{A.}},
\bauthor{\bsnm{Owens}, \binits{M.J.}}:
\byear{2017},
\batitle{The open flux problem}.
\bjtitle{The Astrophysical Journal}
\bvolume{848}(\bissue{1}),
\bfpage{70}.
\doiurl{https://doi.org/10.3847/1538-4357/aa8a70}.
\burl{https://doi.org/10.3847\%2F1538-4357\%2Faa8a70}.
\end{barticle}
\endbibitem

\bibitem[\protect\citeauthoryear{{Liu} \textit{et~al.}}{2012}]{2012liu}
\begin{barticle}
\bauthor{\bsnm{{Liu}}, \binits{Y.}},
\bauthor{\bsnm{{Hoeksema}}, \binits{J.T.}},
\bauthor{\bsnm{{Scherrer}}, \binits{P.H.}},
\bauthor{\bsnm{{Schou}}, \binits{J.}},
\bauthor{\bsnm{{Couvidat}}, \binits{S.}},
\bauthor{\bsnm{{Bush}}, \binits{R.I.}},
\bauthor{\bsnm{{Duvall}}, \binits{T.L.}},
\bauthor{\bsnm{{Hayashi}}, \binits{K.}},
\bauthor{\bsnm{{Sun}}, \binits{X.}},
\bauthor{\bsnm{{Zhao}}, \binits{X.}}:
\byear{2012},
\batitle{{Comparison of Line-of-Sight Magnetograms Taken by the Solar Dynamics
  Observatory/Helioseismic and Magnetic Imager and Solar and Heliospheric
  Observatory/Michelson Doppler Imager}}.
\bjtitle{\solphys}
\bvolume{279}(\bissue{1}),
\bfpage{295}.
\doiurl{https://doi.org/10.1007/s11207-012-9976-x}.
\adsurl{https://ui.adsabs.harvard.edu/abs/2012SoPh..279..295L}.
\end{barticle}
\endbibitem

\bibitem[\protect\citeauthoryear{{Lowder}, {Qiu}, and
  {Leamon}}{2017}]{2017lowder}
\begin{barticle}
\bauthor{\bsnm{{Lowder}}, \binits{C.}},
\bauthor{\bsnm{{Qiu}}, \binits{J.}},
\bauthor{\bsnm{{Leamon}}, \binits{R.}}:
\byear{2017},
\batitle{{Coronal Holes and Open Magnetic Flux over Cycles 23 and 24}}.
\bjtitle{\solphys}
\bvolume{292}(\bissue{1}),
\bfpage{18}.
\doiurl{https://doi.org/10.1007/s11207-016-1041-8}.
\adsurl{https://ui.adsabs.harvard.edu/abs/2017SoPh..292...18L}.
\end{barticle}
\endbibitem

\bibitem[\protect\citeauthoryear{{Lowder} \textit{et~al.}}{2014}]{2014lowder}
\begin{barticle}
\bauthor{\bsnm{{Lowder}}, \binits{C.}},
\bauthor{\bsnm{{Qiu}}, \binits{J.}},
\bauthor{\bsnm{{Leamon}}, \binits{R.}},
\bauthor{\bsnm{{Liu}}, \binits{Y.}}:
\byear{2014},
\batitle{{Measurements of EUV Coronal Holes and Open Magnetic Flux}}.
\bjtitle{\apj}
\bvolume{783},
\bfpage{142}.
\doiurl{https://doi.org/10.1088/0004-637X/783/2/142}.
\adsurl{2014ApJ...783..142L}.
\end{barticle}
\endbibitem

\bibitem[\protect\citeauthoryear{{Ma} \textit{et~al.}}{2014}]{2014ma}
\begin{barticle}
\bauthor{\bsnm{{Ma}}, \binits{L.}},
\bauthor{\bsnm{{Qu}}, \binits{Z.-Q.}},
\bauthor{\bsnm{{Yan}}, \binits{X.-L.}},
\bauthor{\bsnm{{Xue}}, \binits{Z.-K.}}:
\byear{2014},
\batitle{{Interchange reconnection between an active region and a coronal
  hole}}.
\bjtitle{Research in Astronomy and Astrophysics}
\bvolume{14}(\bissue{2}),
\bfpage{221}.
\doiurl{https://doi.org/10.1088/1674-4527/14/2/009}.
\adsurl{https://ui.adsabs.harvard.edu/abs/2014RAA....14..221M}.
\end{barticle}
\endbibitem

\bibitem[\protect\citeauthoryear{{Madjarska} and
  {Wiegelmann}}{2009}]{2009madjarska}
\begin{barticle}
\bauthor{\bsnm{{Madjarska}}, \binits{M.S.}},
\bauthor{\bsnm{{Wiegelmann}}, \binits{T.}}:
\byear{2009},
\batitle{{Coronal hole boundaries evolution at small scales. I. EIT 195 {\AA}
  and TRACE 171 {\AA}view}}.
\bjtitle{\aap}
\bvolume{503},
\bfpage{991}.
\doiurl{https://doi.org/10.1051/0004-6361/200912066}.
\adsurl{https://ui.adsabs.harvard.edu/abs/2009A\%26A...503..991M}.
\end{barticle}
\endbibitem

\bibitem[\protect\citeauthoryear{{Madjarska}, {Doyle}, and {van
  Driel-Gesztelyi}}{2004}]{2004Madjarska}
\begin{barticle}
\bauthor{\bsnm{{Madjarska}}, \binits{M.S.}},
\bauthor{\bsnm{{Doyle}}, \binits{J.G.}},
\bauthor{\bsnm{{van Driel-Gesztelyi}}, \binits{L.}}:
\byear{2004},
\batitle{{Evidence of Magnetic Reconnection along Coronal Hole Boundaries}}.
\bjtitle{\apj}
\bvolume{603}(\bissue{1}),
\bfpage{L57}.
\doiurl{https://doi.org/10.1086/383030}.
\adsurl{https://ui.adsabs.harvard.edu/abs/2004ApJ...603L..57M}.
\end{barticle}
\endbibitem

\bibitem[\protect\citeauthoryear{{Nolte} \textit{et~al.}}{1976}]{1976nolte}
\begin{barticle}
\bauthor{\bsnm{{Nolte}}, \binits{J.T.}},
\bauthor{\bsnm{{Krieger}}, \binits{A.S.}},
\bauthor{\bsnm{{Timothy}}, \binits{A.F.}},
\bauthor{\bsnm{{Gold}}, \binits{R.E.}},
\bauthor{\bsnm{{Roelof}}, \binits{E.C.}},
\bauthor{\bsnm{{Vaiana}}, \binits{G.}},
\bauthor{\bsnm{{Lazarus}}, \binits{A.J.}},
\bauthor{\bsnm{{Sullivan}}, \binits{J.D.}},
\bauthor{\bsnm{{McIntosh}}, \binits{P.S.}}:
\byear{1976},
\batitle{{Coronal holes as sources of solar wind}}.
\bjtitle{\solphys}
\bvolume{46},
\bfpage{303}.
\doiurl{https://doi.org/10.1007/BF00149859}.
\adsurl{1976SoPh...46..303N}.
\end{barticle}
\endbibitem

\bibitem[\protect\citeauthoryear{{Obridko} and {Shelting}}{1989}]{1989Obridko}
\begin{barticle}
\bauthor{\bsnm{{Obridko}}, \binits{V.N.}},
\bauthor{\bsnm{{Shelting}}, \binits{B.D.}}:
\byear{1989},
\batitle{{Coronal holes as indicators of large-scale magnetic fields in the
  corona}}.
\bjtitle{\solphys}
\bvolume{124}(\bissue{1}),
\bfpage{73}.
\doiurl{https://doi.org/10.1007/BF00146520}.
\adsurl{https://ui.adsabs.harvard.edu/abs/1989SoPh..124...73O}.
\end{barticle}
\endbibitem

\bibitem[\protect\citeauthoryear{{Ochsenbein}, {Bauer}, and
  {Marcout}}{2000}]{2000Ochsenbein_vizier}
\begin{barticle}
\bauthor{\bsnm{{Ochsenbein}}, \binits{F.}},
\bauthor{\bsnm{{Bauer}}, \binits{P.}},
\bauthor{\bsnm{{Marcout}}, \binits{J.}}:
\byear{2000},
\batitle{{The VizieR database of astronomical catalogues}}.
\bjtitle{\aaps}
\bvolume{143},
\bfpage{23}.
\doiurl{https://doi.org/10.1051/aas:2000169}.
\adsurl{https://ui.adsabs.harvard.edu/abs/2000A\%26AS..143...23O}.
\end{barticle}
\endbibitem

\bibitem[\protect\citeauthoryear{{Odstr{\v c}il} and {Pizzo}}{1999}]{1999enlil}
\begin{barticle}
\bauthor{\bsnm{{Odstr{\v c}il}}, \binits{D.}},
\bauthor{\bsnm{{Pizzo}}, \binits{V.J.}}:
\byear{1999},
\batitle{{Three-dimensional propagation of CMEs in a structured solar wind
  flow: 1. CME launched within the streamer belt}}.
\bjtitle{\jgr}
\bvolume{104},
\bfpage{483}.
\doiurl{https://doi.org/10.1029/1998JA900019}.
\adsurl{1999JGR...104..483O}.
\end{barticle}
\endbibitem

\bibitem[\protect\citeauthoryear{{Pesnell}, {Thompson}, and
  {Chamberlin}}{2012}]{2012pesnell_SDO}
\begin{barticle}
\bauthor{\bsnm{{Pesnell}}, \binits{W.D.}},
\bauthor{\bsnm{{Thompson}}, \binits{B.J.}},
\bauthor{\bsnm{{Chamberlin}}, \binits{P.C.}}:
\byear{2012},
\batitle{{The Solar Dynamics Observatory (SDO)}}.
\bjtitle{\solphys}
\bvolume{275},
\bfpage{3}.
\doiurl{https://doi.org/10.1007/s11207-011-9841-3}.
\adsurl{2012SoPh..275....3P}.
\end{barticle}
\endbibitem

\bibitem[\protect\citeauthoryear{{Pinto} and {Rouillard}}{2017}]{2017pinto}
\begin{barticle}
\bauthor{\bsnm{{Pinto}}, \binits{R.F.}},
\bauthor{\bsnm{{Rouillard}}, \binits{A.P.}}:
\byear{2017},
\batitle{{A Multiple Flux-tube Solar Wind Model}}.
\bjtitle{\apj}
\bvolume{838},
\bfpage{89}.
\doiurl{https://doi.org/10.3847/1538-4357/aa6398}.
\adsurl{2017ApJ...838...89P}.
\end{barticle}
\endbibitem

\bibitem[\protect\citeauthoryear{{Pomoell} and {Poedts}}{2018}]{2018euhforia}
\begin{barticle}
\bauthor{\bsnm{{Pomoell}}, \binits{J.}},
\bauthor{\bsnm{{Poedts}}, \binits{S.}}:
\byear{2018},
\batitle{{EUHFORIA: European heliospheric forecasting information asset}}.
\bjtitle{Journal of Space Weather and Space Climate}
\bvolume{8}(\bissue{27}),
\bfpage{A35}.
\doiurl{https://doi.org/10.1051/swsc/2018020}.
\adsurl{2018JSWSC...8A..35P}.
\end{barticle}
\endbibitem

\bibitem[\protect\citeauthoryear{{Prato} \textit{et~al.}}{2012}]{2012Prato}
\begin{barticle}
\bauthor{\bsnm{{Prato}}, \binits{M.}},
\bauthor{\bsnm{{Cavicchioli}}, \binits{R.}},
\bauthor{\bsnm{{Zanni}}, \binits{L.}},
\bauthor{\bsnm{{Boccacci}}, \binits{P.}},
\bauthor{\bsnm{{Bertero}}, \binits{M.}}:
\byear{2012},
\batitle{{Efficient deconvolution methods for astronomical imaging: algorithms
  and IDL-GPU codes}}.
\bjtitle{\aap}
\bvolume{539},
\bfpage{A133}.
\doiurl{https://doi.org/10.1051/0004-6361/201118681}.
\adsurl{https://ui.adsabs.harvard.edu/abs/2012A&A...539A.133P}.
\end{barticle}
\endbibitem

\bibitem[\protect\citeauthoryear{{Raymond} and {Doyle}}{1981}]{1981raymond}
\begin{barticle}
\bauthor{\bsnm{{Raymond}}, \binits{J.C.}},
\bauthor{\bsnm{{Doyle}}, \binits{J.G.}}:
\byear{1981},
\batitle{{The energy balance in coronal holes and average quiet-sun regions}}.
\bjtitle{\apj}
\bvolume{247},
\bfpage{686}.
\doiurl{https://doi.org/10.1086/159080}.
\adsurl{https://ui.adsabs.harvard.edu/abs/1981ApJ...247..686R}.
\end{barticle}
\endbibitem

\bibitem[\protect\citeauthoryear{{Reiss} \textit{et~al.}}{2015}]{2015reiss}
\begin{barticle}
\bauthor{\bsnm{{Reiss}}, \binits{M.A.}},
\bauthor{\bsnm{{Hofmeister}}, \binits{S.J.}},
\bauthor{\bsnm{{De Visscher}}, \binits{R.}},
\bauthor{\bsnm{{Temmer}}, \binits{M.}},
\bauthor{\bsnm{{Veronig}}, \binits{A.M.}},
\bauthor{\bsnm{{Delouille}}, \binits{V.}},
\bauthor{\bsnm{{Mampaey}}, \binits{B.}},
\bauthor{\bsnm{{Ahammer}}, \binits{H.}}:
\byear{2015},
\batitle{{Improvements on coronal hole detection in SDO/AIA images using
  supervised classification}}.
\bjtitle{Journal of Space Weather and Space Climate}
\bvolume{5},
\bfpage{A23}.
\doiurl{https://doi.org/10.1051/swsc/2015025}.
\adsurl{https://ui.adsabs.harvard.edu/abs/2015JSWSC...5A..23R}.
\end{barticle}
\endbibitem

\bibitem[\protect\citeauthoryear{{Reiss} \textit{et~al.}}{2016}]{2016reiss}
\begin{barticle}
\bauthor{\bsnm{{Reiss}}, \binits{M.A.}},
\bauthor{\bsnm{{Temmer}}, \binits{M.}},
\bauthor{\bsnm{{Veronig}}, \binits{A.M.}},
\bauthor{\bsnm{{Nikolic}}, \binits{L.}},
\bauthor{\bsnm{{Vennerstrom}}, \binits{S.}},
\bauthor{\bsnm{{Sch{\"o}ngassner}}, \binits{F.}},
\bauthor{\bsnm{{Hofmeister}}, \binits{S.J.}}:
\byear{2016},
\batitle{{Verification of high-speed solar wind stream forecasts using
  operational solar wind models}}.
\bjtitle{Space Weather}
\bvolume{14},
\bfpage{495}.
\doiurl{https://doi.org/10.1002/2016SW001390}.
\adsurl{2016SpWea..14..495R}.
\end{barticle}
\endbibitem

\bibitem[\protect\citeauthoryear{{Reiss} \textit{et~al.}}{2014}]{2014reiss}
\begin{barticle}
\bauthor{\bsnm{{Reiss}}, \binits{M.}},
\bauthor{\bsnm{{Temmer}}, \binits{M.}},
\bauthor{\bsnm{{Rotter}}, \binits{T.}},
\bauthor{\bsnm{{Hofmeister}}, \binits{S.J.}},
\bauthor{\bsnm{{Veronig}}, \binits{A.M.}}:
\byear{2014},
\batitle{{Identification of coronal holes and filament channels in SDO/AIA
  193{\AA} images via geometrical classification methods}}.
\bjtitle{Central European Astrophysical Bulletin}
\bvolume{38},
\bfpage{95}.
\adsurl{2014CEAB...38...95R}.
\end{barticle}
\endbibitem

\bibitem[\protect\citeauthoryear{{Riley}
  \textit{et~al.}}{2012}]{2012riley_CORHEL}
\begin{barticle}
\bauthor{\bsnm{{Riley}}, \binits{P.}},
\bauthor{\bsnm{{Linker}}, \binits{J.A.}},
\bauthor{\bsnm{{Lionello}}, \binits{R.}},
\bauthor{\bsnm{{Mikic}}, \binits{Z.}}:
\byear{2012},
\batitle{{Corotating interaction regions during the recent solar minimum: The
  power and limitations of global MHD modeling}}.
\bjtitle{Journal of Atmospheric and Solar-Terrestrial Physics}
\bvolume{83},
\bfpage{1}.
\doiurl{https://doi.org/10.1016/j.jastp.2011.12.013}.
\adsurl{https://ui.adsabs.harvard.edu/abs/2012JASTP..83....1R}.
\end{barticle}
\endbibitem

\bibitem[\protect\citeauthoryear{{Rotter} \textit{et~al.}}{2012}]{2012rotter}
\begin{barticle}
\bauthor{\bsnm{{Rotter}}, \binits{T.}},
\bauthor{\bsnm{{Veronig}}, \binits{A.M.}},
\bauthor{\bsnm{{Temmer}}, \binits{M.}},
\bauthor{\bsnm{{Vr{\v s}nak}}, \binits{B.}}:
\byear{2012},
\batitle{{Relation Between Coronal Hole Areas on the Sun and the Solar Wind
  Parameters at 1 AU}}.
\bjtitle{\solphys}
\bvolume{281},
\bfpage{793}.
\doiurl{https://doi.org/10.1007/s11207-012-0101-y}.
\adsurl{2012SoPh..281..793R}.
\end{barticle}
\endbibitem

\bibitem[\protect\citeauthoryear{{Rotter} \textit{et~al.}}{2015}]{2015rotter}
\begin{barticle}
\bauthor{\bsnm{{Rotter}}, \binits{T.}},
\bauthor{\bsnm{{Veronig}}, \binits{A.M.}},
\bauthor{\bsnm{{Temmer}}, \binits{M.}},
\bauthor{\bsnm{{Vr{\v{s}}nak}}, \binits{B.}}:
\byear{2015},
\batitle{{Real-Time Solar Wind Prediction Based on SDO/AIA Coronal Hole Data}}.
\bjtitle{\solphys}
\bvolume{290}(\bissue{5}),
\bfpage{1355}.
\doiurl{https://doi.org/10.1007/s11207-015-0680-5}.
\adsurl{https://ui.adsabs.harvard.edu/abs/2015SoPh..290.1355R}.
\end{barticle}
\endbibitem

\bibitem[\protect\citeauthoryear{{Scherrer}
  \textit{et~al.}}{1995}]{1995scherrer_mdi}
\begin{barticle}
\bauthor{\bsnm{{Scherrer}}, \binits{P.H.}},
\bauthor{\bsnm{{Bogart}}, \binits{R.S.}},
\bauthor{\bsnm{{Bush}}, \binits{R.I.}},
\bauthor{\bsnm{{Hoeksema}}, \binits{J.T.}},
\bauthor{\bsnm{{Kosovichev}}, \binits{A.G.}},
\bauthor{\bsnm{{Schou}}, \binits{J.}},
\bauthor{\bsnm{{Rosenberg}}, \binits{W.}},
\bauthor{\bsnm{{Springer}}, \binits{L.}},
\bauthor{\bsnm{{Tarbell}}, \binits{T.D.}},
\bauthor{\bsnm{{Title}}, \binits{A.}},
\bauthor{\bsnm{{Wolfson}}, \binits{C.J.}},
\bauthor{\bsnm{{Zayer}}, \binits{I.}},
\bauthor{\bsnm{{MDI Engineering Team}}}:
\byear{1995},
\batitle{{The Solar Oscillations Investigation - Michelson Doppler Imager}}.
\bjtitle{\solphys}
\bvolume{162},
\bfpage{129}.
\doiurl{https://doi.org/10.1007/BF00733429}.
\adsurl{https://ui.adsabs.harvard.edu/abs/1995SoPh..162..129S}.
\end{barticle}
\endbibitem

\bibitem[\protect\citeauthoryear{{Schou} \textit{et~al.}}{2012}]{2012schou_HMI}
\begin{barticle}
\bauthor{\bsnm{{Schou}}, \binits{J.}},
\bauthor{\bsnm{{Scherrer}}, \binits{P.H.}},
\bauthor{\bsnm{{Bush}}, \binits{R.I.}},
\bauthor{\bsnm{{Wachter}}, \binits{R.}},
\bauthor{\bsnm{{Couvidat}}, \binits{S.}},
\bauthor{\bsnm{{Rabello-Soares}}, \binits{M.C.}},
\bauthor{\bsnm{{Bogart}}, \binits{R.S.}},
\bauthor{\bsnm{{Hoeksema}}, \binits{J.T.}},
\bauthor{\bsnm{{Liu}}, \binits{Y.}},
\bauthor{\bsnm{{Duvall}}, \binits{T.L.}},
\bauthor{\bsnm{{Akin}}, \binits{D.J.}},
\bauthor{\bsnm{{Allard}}, \binits{B.A.}},
\bauthor{\bsnm{{Miles}}, \binits{J.W.}},
\bauthor{\bsnm{{Rairden}}, \binits{R.}},
\bauthor{\bsnm{{Shine}}, \binits{R.A.}},
\bauthor{\bsnm{{Tarbell}}, \binits{T.D.}},
\bauthor{\bsnm{{Title}}, \binits{A.M.}},
\bauthor{\bsnm{{Wolfson}}, \binits{C.J.}},
\bauthor{\bsnm{{Elmore}}, \binits{D.F.}},
\bauthor{\bsnm{{Norton}}, \binits{A.A.}},
\bauthor{\bsnm{{Tomczyk}}, \binits{S.}}:
\byear{2012},
\batitle{{Design and Ground Calibration of the Helioseismic and Magnetic Imager
  (HMI) Instrument on the Solar Dynamics Observatory (SDO)}}.
\bjtitle{\solphys}
\bvolume{275},
\bfpage{229}.
\doiurl{https://doi.org/10.1007/s11207-011-9842-2}.
\adsurl{2012SoPh..275..229S}.
\end{barticle}
\endbibitem

\bibitem[\protect\citeauthoryear{{Schwenn}}{2006}]{schwenn06}
\begin{barticle}
\bauthor{\bsnm{{Schwenn}}, \binits{R.}}:
\byear{2006},
\batitle{{Solar Wind Sources and Their Variations Over the Solar Cycle}}.
\bjtitle{\ssr}
\bvolume{124},
\bfpage{51}.
\doiurl{https://doi.org/10.1007/s11214-006-9099-5}.
\adsurl{2006SSRv..124...51S}.
\end{barticle}
\endbibitem

\bibitem[\protect\citeauthoryear{{Temmer}, {Hinterreiter}, and
  {Reiss}}{2018}]{temmer18}
\begin{barticle}
\bauthor{\bsnm{{Temmer}}, \binits{M.}},
\bauthor{\bsnm{{Hinterreiter}}, \binits{J.}},
\bauthor{\bsnm{{Reiss}}, \binits{M.A.}}:
\byear{2018},
\batitle{Coronal hole evolution from multi-viewpoint data as input for a stereo
  solar wind speed persistence model}.
\bjtitle{J. Space Weather Space Clim.}
\bvolume{8},
\bfpage{A18}.
\doiurl{https://doi.org/10.1051/swsc/2018007}.
\burl{https://doi.org/10.1051/swsc/2018007}.
\end{barticle}
\endbibitem

\bibitem[\protect\citeauthoryear{{Tokumaru}
  \textit{et~al.}}{2017}]{2017tokumaru}
\begin{barticle}
\bauthor{\bsnm{{Tokumaru}}, \binits{M.}},
\bauthor{\bsnm{{Satonaka}}, \binits{D.}},
\bauthor{\bsnm{{Fujiki}}, \binits{K.}},
\bauthor{\bsnm{{Hayashi}}, \binits{K.}},
\bauthor{\bsnm{{Hakamada}}, \binits{K.}}:
\byear{2017},
\batitle{{Relation Between Coronal Hole Areas and Solar Wind Speeds Derived
  from Interplanetary Scintillation Measurements}}.
\bjtitle{\solphys}
\bvolume{292},
\bfpage{41}.
\doiurl{https://doi.org/10.1007/s11207-017-1066-7}.
\adsurl{2017SoPh..292...41T}.
\end{barticle}
\endbibitem

\bibitem[\protect\citeauthoryear{{Verbeeck} \textit{et~al.}}{2014}]{2014spoca}
\begin{barticle}
\bauthor{\bsnm{{Verbeeck}}, \binits{C.}},
\bauthor{\bsnm{{Delouille}}, \binits{V.}},
\bauthor{\bsnm{{Mampaey}}, \binits{B.}},
\bauthor{\bsnm{{De Visscher}}, \binits{R.}}:
\byear{2014},
\batitle{{The SPoCA-suite: Software for extraction, characterization, and
  tracking of active regions and coronal holes on EUV images}}.
\bjtitle{\aap}
\bvolume{561},
\bfpage{A29}.
\doiurl{https://doi.org/10.1051/0004-6361/201321243}.
\adsurl{2014A\%26A...561A..29V}.
\end{barticle}
\endbibitem

\bibitem[\protect\citeauthoryear{{Vr{\v s}nak}, {Temmer}, and
  {Veronig}}{2007}]{2007vrsnak}
\begin{barticle}
\bauthor{\bsnm{{Vr{\v s}nak}}, \binits{B.}},
\bauthor{\bsnm{{Temmer}}, \binits{M.}},
\bauthor{\bsnm{{Veronig}}, \binits{A.M.}}:
\byear{2007},
\batitle{{Coronal Holes and Solar Wind High-Speed Streams: I. Forecasting the
  Solar Wind Parameters}}.
\bjtitle{\solphys}
\bvolume{240},
\bfpage{315}.
\doiurl{https://doi.org/10.1007/s11207-007-0285-8}.
\adsurl{2007SoPh..240..315V}.
\end{barticle}
\endbibitem

\bibitem[\protect\citeauthoryear{{Wallace} \textit{et~al.}}{2019}]{2019wallace}
\begin{barticle}
\bauthor{\bsnm{{Wallace}}, \binits{S.}},
\bauthor{\bsnm{{Arge}}, \binits{C.N.}},
\bauthor{\bsnm{{Pattichis}}, \binits{M.}},
\bauthor{\bsnm{{Hock-Mysliwiec}}, \binits{R.A.}},
\bauthor{\bsnm{{Henney}}, \binits{C.J.}}:
\byear{2019},
\batitle{{Estimating Total Open Heliospheric Magnetic Flux}}.
\bjtitle{\solphys}
\bvolume{294},
\bfpage{19}.
\doiurl{https://doi.org/10.1007/s11207-019-1402-1}.
\adsurl{2019SoPh..294...19W}.
\end{barticle}
\endbibitem

\bibitem[\protect\citeauthoryear{{Wang} and {Sheeley}}{2004}]{2004wang}
\begin{barticle}
\bauthor{\bsnm{{Wang}}, \binits{Y.-M.}},
\bauthor{\bsnm{{Sheeley}}, \binits{J.} \bsuffix{N.~R.}}:
\byear{2004},
\batitle{{Footpoint Switching and the Evolution of Coronal Holes}}.
\bjtitle{\apj}
\bvolume{612}(\bissue{2}),
\bfpage{1196}.
\doiurl{https://doi.org/10.1086/422711}.
\adsurl{https://ui.adsabs.harvard.edu/abs/2004ApJ...612.1196W}.
\end{barticle}
\endbibitem

\bibitem[\protect\citeauthoryear{{Wendeln} and {Landi}}{2018}]{2018Wendeln}
\begin{barticle}
\bauthor{\bsnm{{Wendeln}}, \binits{C.}},
\bauthor{\bsnm{{Landi}}, \binits{E.}}:
\byear{2018},
\batitle{{EUV Emission and Scattered Light Diagnostics of Equatorial Coronal
  Holes as Seen by Hinode/EIS}}.
\bjtitle{\apj}
\bvolume{856},
\bfpage{28}.
\doiurl{https://doi.org/10.3847/1538-4357/aaaadf}.
\adsurl{https://ui.adsabs.harvard.edu/abs/2018ApJ...856...28W}.
\end{barticle}
\endbibitem

\bibitem[\protect\citeauthoryear{{Yang} \textit{et~al.}}{2011}]{2011yang}
\begin{barticle}
\bauthor{\bsnm{{Yang}}, \binits{S.}},
\bauthor{\bsnm{{Zhang}}, \binits{J.}},
\bauthor{\bsnm{{Li}}, \binits{T.}},
\bauthor{\bsnm{{Liu}}, \binits{Y.}}:
\byear{2011},
\batitle{{SDO Observations of Magnetic Reconnection At Coronal Hole
  Boundaries}}.
\bjtitle{\apjl}
\bvolume{732},
\bfpage{L7}.
\doiurl{https://doi.org/10.1088/2041-8205/732/1/L7}.
\adsurl{https://ui.adsabs.harvard.edu/abs/2011ApJ...732L...7Y}.
\end{barticle}
\endbibitem

\end{thebibliography}
%
%
%
%

\end{article} 
\end{document}